\documentclass[aip,jcp,preprint,onecolumn,floatfix]{revtex4-1}

\usepackage{graphicx}
\usepackage{color}
\usepackage{amsmath}
\usepackage{lineno}
\usepackage{float}

\def\figref#1{Fig.~\ref{#1}}
\def\eqnref#1{Eq.~(\ref{#1})}
\def\secref#1{Sect.~\ref{#1}}

\begin{document}
	\title{Evidence for multiple scattering effects in the electron mobility in dense argon gas} 
	\author{A. F. Borghesani}
	\email[Corresponding author: ]{armandofrancesco.borghesani@unipd.it}
	\affiliation{Dipartimento di Fisica ed Astronomia, Universit{\`a} degli Studi, Padova, Italy} 
	\author{P. Lamp}
	\affiliation{
MPI f{\H u}r Physik, Garching, Germany}
	\altaffiliation{Present address: Scires Battery Technologies GmbH,
		D-80331 Munich}
	\begin{abstract}
		We report 
		measurements of the electron drift mobility in dense argon gas over an extended range of densities, temperatures, and electric fields, supplementing our earlier work. The new measurements confirm the validity of the heuristic model we previously developed by introducing multiple scattering effects in the classical kinetic theory description of the electron mobility in a dilute gas. We definitively show that, in the argon gas, because of the particular energy dependence of its electron-atom momentum-transfer scattering cross section, none of the multiple scattering effects we have identified in the past can be neglected if the mobility behavior is to be accurately rationalized over the whole investigated parameter range.
	\end{abstract}	
	\maketitle 
	\section{Introduction}\label{sec:intro}
	The investigation of the transport properties of low-energy quasi-free electrons in dense non-polar gases has long been pursued due to its technological relevance and because interesting insights can be gained into the interaction of an excess electron in a disordered medium and its dynamics and energetics.  The delicate interplay between the density of the environment and the quantum nature of the electron gives rise to multiple scattering effects (MSE) that affect the dependence of electron transport on gas parameters, such as temperature and density~\cite{OMa80}.  
	
	Under such conditions, the fundamental assumption of the classical kinetic theory (CKT)~\cite{hux1974} is no longer valid, that is, that the interaction between the excess electrons and the gas atoms occurs via a series of well-defined successive binary collisions. As a consequence, when electrons drift through the gas under the action of an externally applied electric field \(E,\) the  CKT  prediction for the zero-field, density-normalized electron drift mobility \(\mu_0 N,\) i.e., 
	\begin{equation}
		\mu_0 N = \frac{4e}{3\left(2\pi m\right)^{1/2}\left(k_\mathrm{B}T\right)^{5/2}}\int\limits_0^\infty\frac{\varepsilon}{\sigma_\mathrm{mt}(\varepsilon)}\mathrm{e}^{-\varepsilon/k_\mathrm{B}T}\,\mathrm{d}\varepsilon
		\label{eq:mu0Ncl}
	\end{equation}
	fails to describe the experimentally observed unexpected density dependence of \(\mu_0 N .\) Here, \(N\) is the gas number density, \(e\) and \(m\) are the electric charge and mass of the electron, respectively. \(T\) is the temperature, \(k_\mathrm{B}\) is the Boltzmann constant, \(\varepsilon\) is the electron energy, and \(\sigma_\mathrm{mt}\) is the energy-dependent electron-atom momentum-transfer scattering cross section.
	
	Anomalous density effects, i.e., density-dependent deviations of \(\mu_0 N\) from the classical prediction~\eqnref{eq:mu0Ncl}, have been previously discovered  not only in noble gases~\cite{Gru68,Bartels1975,Bor1985}, which are the simplest possible examples of a disordered system, but also in molecular gases~\cite{Lehning68,Krebs96,Krebs96b}. In gases such as helium and neon (also referred to as {\em repulsive gases}), whose interaction with an excess electron is dominated by the short-range repulsive exchange forces leading to a positive scattering length, the drift mobility exhibits a negative density effect, that is, \(\mu_0 N\) decreases with increasing \(N.\)  In heavier noble gases (also referred to as {\em attractive gases}), such as argon, in which the electron-atom scattering process is dominated by the long-range attractive polarization interaction that gives rise to  a negative scattering length, a positive density effect  is observed and \(\mu_0 N\) increases with increasing \(N\).
	
	It was long realized that the relative magnitudes of the electron quantum wavelength, of the average interatomic distance, and of the electron mean free path are responsible for the appearance of multiple scattering effects. Although it was recognized that they  give rise to a quantum density dependent shift of the electron energy and to the emergence of a mobility edge, different theories for \(\mu_0 N\) were nonetheless proposed depending on the sign of the scattering length so that the two different anomalous density effects were not explained within a unified theoretical framework~\cite{Atr1977,Braglia1982,OMa1992}. 
	
	As we believe that the description of the physical process of scattering of thermal electrons by noble gas atoms at high density should be unique and independent of the sign of the scattering length, we previously proposed a model that identifies three main MSE, which are heuristically incorporated in the CKT to describe the drift mobility \(\mu\) of low-energy electrons in a dense gas~\cite{Bor92}.   
	
	These three MSE are 
	\begin{itemize}
		\item [i -- ] the density-dependent quantum shift of the electron kinetic energy;
		\item[ii -- ] the correlation among scatterers, related to the gas compressibility;
		\item[iii -- ] the  enhancement of the electron backscattering rate  due to quantum self-interference of the electron wave packet scattered off atoms located along paths connected by time-reversal symmetry (eventually leading to the presence of a mobility edge if the cross section is large).
	\end{itemize}
	All these three effects must be taken into account simultaneously, although their individual relevance depends on several factors, among which the most important are the size and energy dependence of the electron-atom scattering cross section, and the thermodynamic state of the gas. 
	
	Our heuristic model has the advantage of retaining the classical picture of binary collisions while introducing an effective density-dependent scattering cross section obtained by suitably ``dressing'' the bare electron-atom scattering cross section with the MSE.
	
	The first of these is related to the fact that the electron energy at the bottom of the conduction band in the gas is shifted by a density-dependent amount~\cite{Spri68} 
	\begin{equation}
		V_0(N)=U_P(N) + E_k(N)\label{eq:v0}
	\end{equation}
	{\em En passant}, we note that the total energy shift \(V_0(N)\) is a highly relevant physical property of quasi-free (or, nearly quasi-free) electrons in dense gaseous environments if several additional phenomena not directly related  to electron transport-- such as the resonant electron attachment to oxygen molecules~\cite{Ner97,Borghesani2020} or the infrared emission of xenon excimers in dense gases~\cite{Borghesani2001IR,Borghesani2025}-- are to be rationalized.   
	
	\(U_P\) is a negative potential energy term that arises from the screened polarization interaction of the electron with the atoms of the gas. 
	\(E_k\) is a kinetic energy contribution  due to the exclusion of the electron from the hard-core volume of the atoms. It can be explicitly computed by replacing the gas structure with an ordered array of hard-sphere scatterers that is locally invariant under translations of amplitude \(2R_\mathrm{WS},\) where \(R_\mathrm{WS}\) is the radius of the Wigner-Seitz (WS) sphere centered on each atom that is defined in terms of the  density \(N\) as \((4\pi/3) R_\mathrm{WS}^3N=1.\)
	
	In order to comply with the translational symmetry requirements, the electron \(s\)-wave function must be antisymmetric  and vanish at \(r=R_\mathrm{WS}\), thereby leading to the eigenvalue equation~\cite{Her91}
	\begin{equation}
		\tan{\left[k_0\left(R_\mathrm{WS}+\frac{\eta_0(k_0)}{k_0}\right)\right]}=k_0R_\mathrm{WS}\label{eq:evko}
	\end{equation}
	Here, 
	\(k_0(N)\) is the ground state electron wave vector and \(\eta_0\) is the \(s\)-wave phaseshift~\cite{OMa63,Her91}. To take into account the superposition of the atomic potential tails \(-\eta_0/k_0\) is replaced by the hard-core radius of the Hartree-Fock potential~\cite{Spri68} \(a=\sqrt{\sigma_T(k_0)/4\pi},\) in which \(\sigma_T\) is the total scattering cross section. Finally, we get
	\begin{equation}
		E_k(N)= \frac{\hbar^2k_0^2(N)}{2m}
		\label{eq:ekk0}
	\end{equation}
	Here, \(\hbar=h/2\pi\) is the reduced Planck constant.
	The scattering processes depend on the electron kinetic energy \(\varepsilon\), which is thus shifted to \(\varepsilon^\prime =\varepsilon +E_k(N)\), in correspondence to which dynamic quantities such as cross sections are to be evaluated. It is evident that this first multiple scattering effect is important only if the cross section is a rapidly varying function of the electron kinetic energy.
	
	The second multiple scattering effect is the correlation among scatterers due to the fact that the electron wave packet spans a region of linear size of the order of its quantum wavelength \(\lambda=\hbar/\sqrt{2m\varepsilon}\) that may encompass several atoms. The electron is then scattered by them simultaneously and the total amplitude of the scattered wave is the coherent sum of all of the partial scattering amplitudes contributed by each atom. As a result, the cross section is enhanced by the static structure factor of the gas~\cite{Lekner1968}, whose long-wavelength limit is \(S(0)=Nk_\mathrm{B}T\chi_T\), in which  \(\chi_T\) is the compressibility of the isothermal gas. This effect can be very relevant if measurements are carried out close to the critical point of liquid-vapor of the gas, where \(\chi_T\) can be very large.
	
	The third multiple scattering effect is an increase in the backscattering rate due to quantum self-interference of the electron wave function scattered off atoms positioned along paths connected by time-reversal symmetry~\cite{Asca1992}. The driving parameter is the ratio of 
	\(\lambda\) to the mean free path of the electron \(\ell =1/N\sigma_\mathrm{mt}.\) In the present case, the drift mobility \(\mu\) is quite large and so is \(\ell,\) whereas the temperatures are not low enough for electrons to endow themselves with a very long thermal wavelength. 
	Thus, a perturbative approximation is sufficient and a linearized form of the scattering rate \(\nu(\varepsilon)\) is readily obtained as
	\begin{equation}
		\nu(\varepsilon)=\nu_0(\varepsilon)\left(
		1+f\frac{\lambda}{\ell}
		\right)\label{eq:nu}
	\end{equation}
	in which \(\nu_0=\sqrt{2\varepsilon/m}N\sigma_\mathrm{mt}(\varepsilon)\)
	is the scattering rate in the dilute-gas limit and \(f\) is a number of order unity~\cite{Atr1977,Polischuk1984}.
	
	These MSE are now incorporated into the CKT equations leading to the formulation of our heuristic model in a way that does not need any adjustable parameters.
	The density-normalized mobility \(\mu N\) is derived from the CKT as
	\begin{equation}
		\mu N =-\frac{2}{3}\left(\frac{e}{m}\right)^{1/2} \int\limits_0^\infty
		\frac{\varepsilon}{\sigma^\star_\mathrm{mt}(\varepsilon)}\frac{\mathrm{d}g(\varepsilon)}{\mathrm{d\varepsilon}}\,\mathrm{d}\varepsilon
		\label{eq:muN}
	\end{equation}
	in which  \(g(\varepsilon)\) is the Davidov-Pidduck electron energy distribution function~\cite{hux1974,Coh67}
	\begin{equation}
		g(\varepsilon)=A\exp{\left\{-
			\int\limits_0^\varepsilon\left[
			k_\mathrm{B}T+\frac{M(eE/N)^2}{6mz{\sigma^\star}^2_\mathrm{mt}(z)} 
			\right]^{-1}\,\mathrm{d}z
			\right\}}\label{eq:DavPid}
	\end{equation}
	Here, \(M\) is the atomic mass, and 
	\(A\) is obtained by enforcing  normalization as \(\int_0^\infty \varepsilon^{1/2}g(\varepsilon)\,\mathrm{d}\varepsilon=1.\)
	\(\sigma_\mathrm{mt}^\star(\varepsilon)\) is the effective momentum transfer scattering cross section that takes into account the MSE and is given by~\cite{Bor92}
	\begin{equation}
		\sigma_\mathrm{mt}^\star(\varepsilon)=\mathcal{F}(w)\sigma_\mathrm{mt}(w)\left[
		1-f \hbar \frac{\mathcal{F}(w)N\sigma_\mathrm{mt}(w)}{\left(2mw\right)^{1/2}}
		\right]^{-1}
		\label{eq:effsmt}
	\end{equation}
	in which \(w=\varepsilon + E_k(N)\) is the shifted energy, and \(\sigma_\mathrm{mt}\) is the dilute-gas limit of the electron-atom momentum transfer scattering cross section.
	The factor \(\mathcal{F}\) takes into account the correlation among scatterers as
	\begin{equation}
		\mathcal{F}(w) = \frac{1}{4w^2}\int\limits_0^{2w} q^3S(q) \,\mathrm{d}q
		\label{eq:F}
	\end{equation}
	The static structure factor \(S(q)\) can be expressed for a not too large exchanged momentum \(\hbar q\) within the Ornstein-Zernike approximation as~\cite{Sta} 
	\begin{equation}
		S(q)= \frac{S(0)+(qL)^2}{1+(qL)^2}
		\label{eq:SK}
	\end{equation}
	The length \(L\) has been determined by X-ray scattering experiments in the form \(L^2\approx 0.1l^2[S(0)-1]\) with the short-range correlation length \(l\) of the order   \(0.5\,\mbox{to}\, 1\,\)nm\cite{Tho63}.
	
	Once more, we remark that only the knowledge of the electron-atom cross section and of the equation of state of the gas is needed to work out the heuristic model,  and no adjustable parameters are necessary. 
	
	The heuristic model explains very successfully the behavior of \(\mu_0 N\) in dense neon~\cite{Bor88,Bor90b} and helium~\cite{Bor21} gases for densities below the onset of electron self-localization in bubbles~\cite{Borghesani1992b,Borghesani2002,Bor21} and recently also in H\(_2\) gas~\cite{Bor25}. In argon, we successfully adopted it to explain the measurements at \(T\approx 162.7\,\)K~\cite{Bor92} and, partly, those closer to the critical temperature at \(T= 152.15\,\)K~\cite{Bor2001,Borghesani2001}. 
	
	Thus, we have carried out measurements in argon gas at several more temperatures to further confirm the model. 
	
	In~\secref{sec:expdet} we briefly describe the experimental technique and apparatus. In~\secref{sec:expres} the results are presented and discussed. Finally, in \secref{sec:conc} we draw some conclusions.

	\section{Experimental details}\label{sec:expdet}
	The measurements were mostly carried out at the Max-Planck-Institut in Munich (Germany) and partly at the Department of Physics of the Padua University in Padua (Italy), namely those at \(T= 152.15\,\)K and \(T=162.3\,\)K.
	The two experimental setups were very similar and we refer to the literature for more technical details~\cite{Eib90,Bor92,Bor2001}. We recall here only their main features.
	
	Both experiments are based on the well-known Townsend pulsed photoemission technique. A bunch of electrons is released from a photocathode irradiated by a short pulse of a Xenon flashlamp~\cite{Bor86} and are drifted to a  collecting anode over a drift distance \(d\) by the action of an externally applied electric field \(E\). The amount of injected charge roughly ranges between 4 and 1000 fC, depending on the gas density and the electric field strength, and is small enough to avoid space-charge effects~\cite{Bor90}.
	
	The current signal produced by drifting electrons is integrated to improve the signal-to-noise ratio~\cite{Bor90}. To further improve it, several tens of electronic signals are acquired and averaged together for each setting of the experimental parameters. Offline analysis of the acquired signals leads to the determination of the drift time \(\tau_e\) and to the computation of the electron drift mobility \(\mu=d/\tau E\)~\cite{Bor90,Bor90c,Eib90}. The typical experimental accuracy is estimated to be \(\Delta \mu/\mu\approx 5\,\) to \(10\,\%.\)
	
	We used ultra-high purity argon gas (N60 grade) with a nominal oxygen content \(1\,\)parts-per-million (p.p.m.). 
	In order to have electronic signals that are not spoiled by electron attachment to molecular oxygen, the gas needs to be further purified. To this end, similar gas purification schemes are adopted in the two experiments~\cite{Eib90,Torzo90,Bor90c}. In both cases, we estimate an oxygen concentration in the range of a fraction of part per billion (p.p.b.)~\cite{Ada88}.
	
	Also, the temperature control is very similar in the two experimental setups. The pressure cell is mounted within a triple-shield thermostat and cooled by using a commercial cryocooler. The temperature is controlled by means of PID controllers, commercial~\cite{Eib90,Bor92} or home-made~\cite{Bru85,Bru86}. In both cases, the temperature is controlled to within \(\Delta T=  0.01\,\)K, or better. 
	
	The gas pressure is measured with an accuracy \(\Delta P\approx 1\,\)kPa and the gas density \(N\) is calculated by exploiting the equation of state available in the literature~\cite{Wag99}.
	
	\section{Experimental results and discussion}\label{sec:expres}
	
	We report here the new measurements of the drift mobility of quasi-free electrons in wide temperature, density, and electric field ranges. These are supplementing (and confirming) previous data obtained at \(T\approx 162.7\,\)K~\cite{Bor92}, \(T\approx 152.15,\,\)K~\cite{Bor2001}, and \(T=162.3\,\)K~\cite{Bor2001}.
	\subsection{Reduced electric field dependence}\label{sec:munvsen}
\begin{figure}[t!]
		\centering\includegraphics[width=\columnwidth]{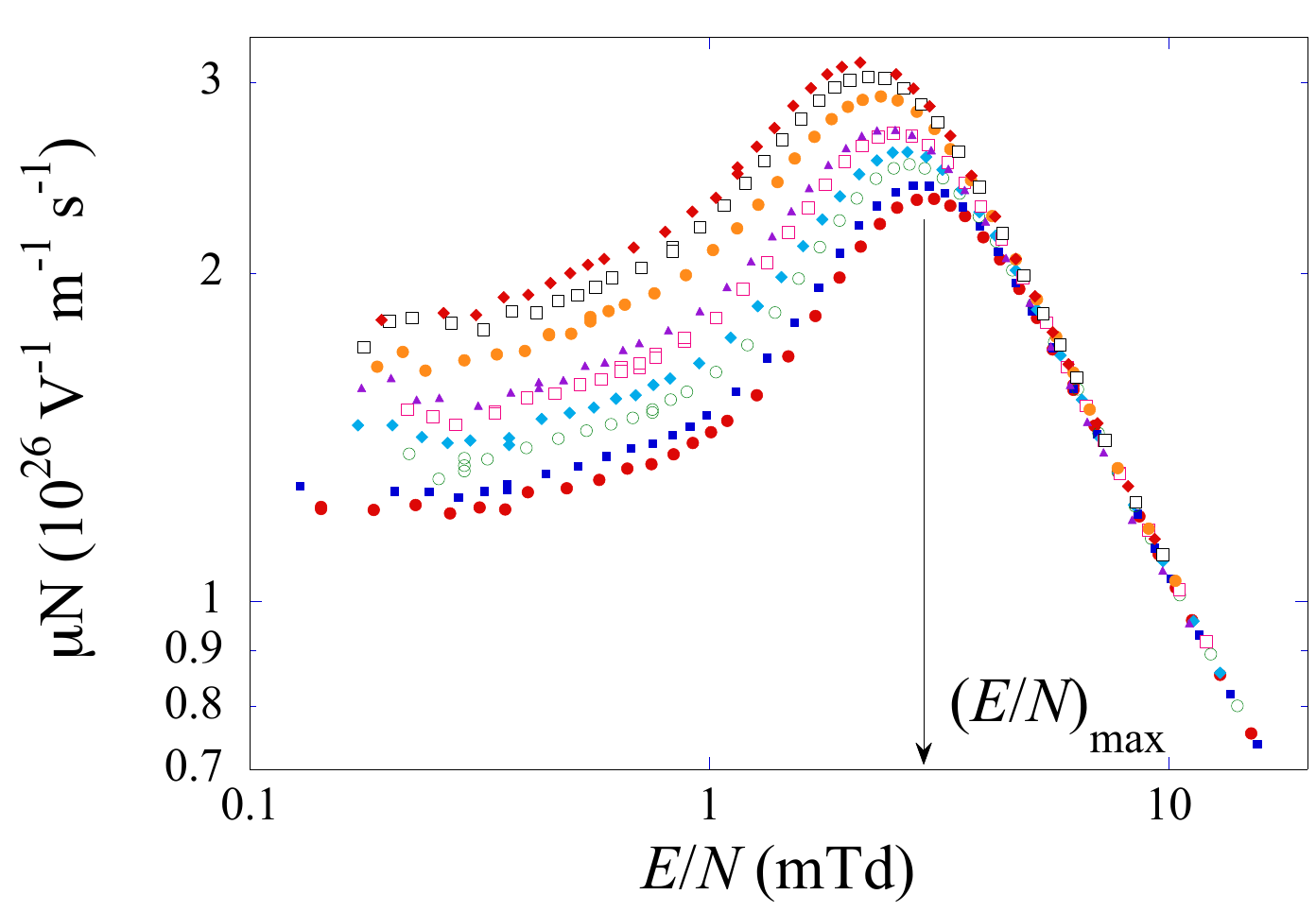}
		\caption{\small Density normalized mobility \(\mu N\) as a function of the reduced electric field \(E/N\) at \(T=177.3\,\)K for several densities \(N\).  From top: \(N\, (10^{26}\, \mbox{m}^{-3}) =\,42.9,\,41.3,\,38.6,\,36.0,\,33.2,\,30.9,\,28.3,\,25.7,\,23.1. \)  The arrow indicates the reduced electric field strength \((E/N)_\mathrm{max}\), at which \(\mu N\) is maximum. (\(1\,\)mTd\(\, = 10^{-24}\,\mbox{Vm}^2\)).\label{fig:figure1}}
	\end{figure}

\begin{figure}[b!] \centering\includegraphics[width=\columnwidth]{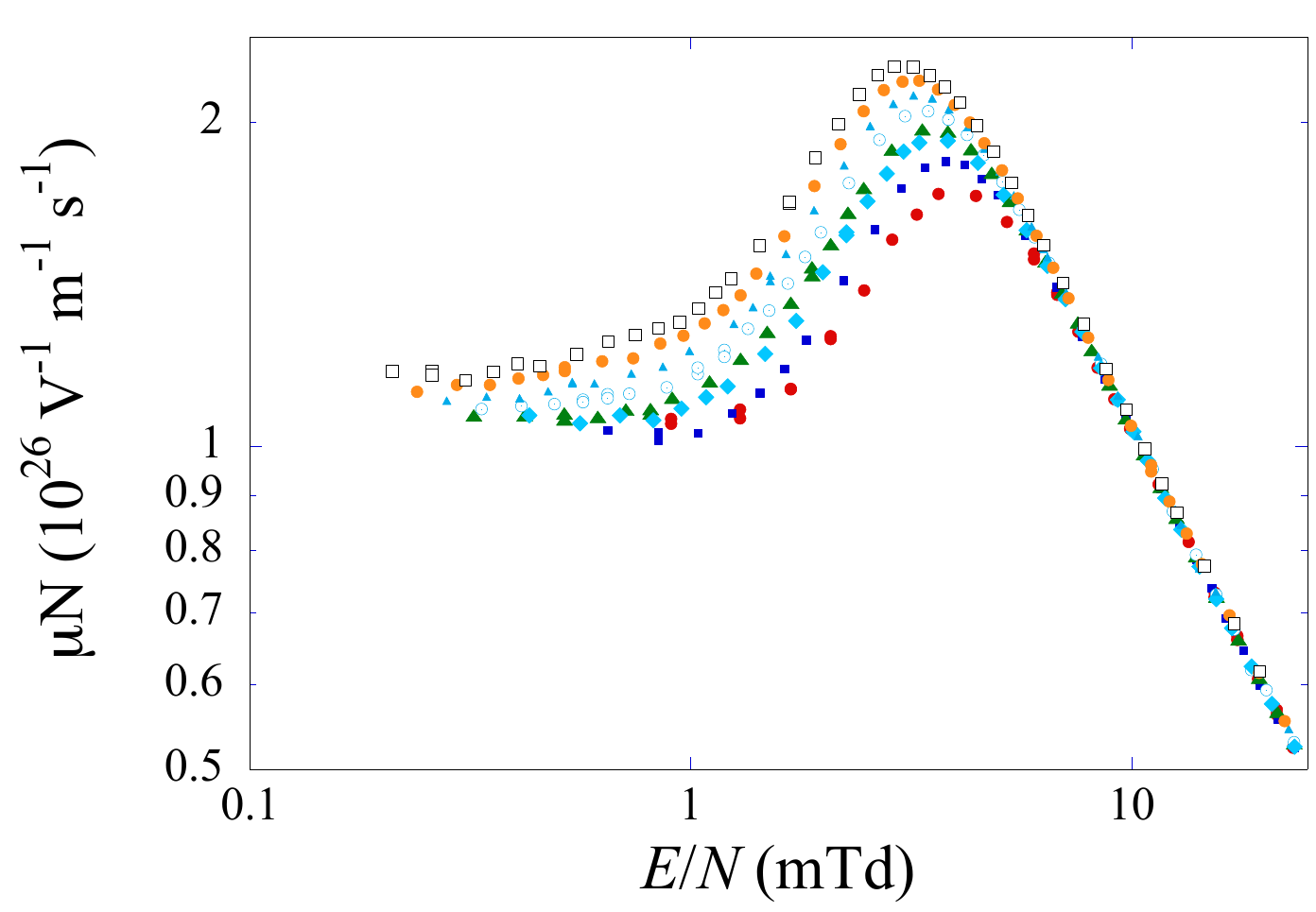}
		\caption{\small Density normalized mobility \(\mu N\) as a function of the reduced electric field \(E/N\) at \(T=177.3\,\)K for several densities \(N\).  From top: \(N\, (10^{26}\, \mbox{m}^{-3}) =\,20.5,\, 18.0,\,15.4,\, 12.8,\, 10.3, \, 7.67, \, 5.10,\, 2.55. \)  (\(1\,\)mTd\(\, = 10^{-24}\,\mbox{Vm}^2\)).\label{fig:figure2}}
	\end{figure}

	As an example, in~\figref{fig:figure1}  and in ~\figref{fig:figure2} we show the dependence of the density-normalized mobility \(\mu N\) on the reduced field \(E/N\) for the densities investigated at \(T=177.3\,\)K. 
	We also carried out measurements at \(T=199.6\,\)K, \(T=152.7\,\)K, and \(T=142.6\,\)K. The last temperature is below the critical one, \(T_c\approx 150.86\,\)K.  
	
	At all investigated temperatures \(\mu N\) shows the same behaviour as a function of the reduced field \(E/N\). For all densities, \(\mu N\) is constant at low \(E/N\), where electrons are still in thermal equilibrium with the gas. As \(E/N\) increases, electrons are heated by the field and become epithermal. As the momentum-transfer scattering cross section \(\sigma_\mathrm{mt}\) rapidly decreases with increasing electron energy, as shown in~\figref{fig:figure3}, \(\mu N\) also increases with \(E/N,\). The presence of the Ramsauer-Townsend minimum in \(\sigma_\mathrm{mt}\) at  energy \(\varepsilon_\mathrm{RT}\) produces the mobility maximum that occurs for the reduced electric field strength \((E/N)_\mathrm{max}\). We infer that the presence of the mobility maximum is a strong indication of the value of the average electron energy at \((E/N)_\mathrm{max}.\) 
	For higher values of \(E/N\), the \(\mu N\) curves corresponding to different density values merge to a single common curve.  For such large values of \(E/N\) the kinetic energy shift \(E_k(N)\) is negligible with respect to the energy gained by the electrons from the field so that the multiple scattering effects become irrelevant.
	\mbox{}
	These new measurements confirm the result of the previous ones~\cite{Bor92,Bor2001}. In particular, in contrast to the prediction of the CKT, \(\mu N\) also shows a dependence on the gas density \(N\), especially  in the zero-field limit. Actually, the zero-field density normalized mobility \(\mu_0 N\) increases with increasing \(N, \) as easily noticed by inspecting~\figref{fig:figure1} and~\figref{fig:figure2}. Moreover, we can also realize that \((E/N)_\mathrm{max}\) decreases with increasing \(N.\) These two observed behaviors will be easily rationalized by invoking the kinetic energy shift.
	
	We will now show that, in accordance with our previous measurements, all these experimental features can be explained by accounting for the effect of the three multiple scattering mechanisms that we have identified in the past in noble gases~\cite{Bor88,Bor90b,Bor21} and that we have also  observed in the H\(_2\) molecular gas~\cite{Bor25}.
	
	\subsection{Density dependence of the zero-field density normalized mobility}\label{sec:mu0nvsN}
	As stated in~\secref{sec:intro}, the three multiple scattering effects are: i) the density-dependent quantum shift \(E_k(N)\) of the electron kinetic energy at the bottom of the conduction band in the gas, ii) the correlation among scatterers, and  iii) the quantum self-interference of the electron wave packet scattering off atoms located along paths connected by time-reversal symmetry.
	The relative weight of the three multiple scattering effects depends on the features of the momentum-transfer scattering cross section and on the thermodynamic state of the gas. 
	
	For large and almost energy independent cross sections, as is the case of helium and hydrogen, and not very close to the critical point, the kinetic energy shift \(E_k(N)\) and the correlation among scatterers can be safely neglected, whereas the quantum self-interference effect determines almost uniquely the density dependence of \(\mu_0 N\)~\cite{Bor21,Bor25}.
	
	In contrast, in gases such as neon, whose cross section is rather small but varies very rapidly with the electron energy, the dominant effect is due mainly to the kinetic energy shift, whereas the two other MSE give negligible contribution to the observed mobility~\cite{Bor88,Bor90b}.
		\begin{figure}[t!]
		\centering
		\includegraphics[width=\columnwidth]{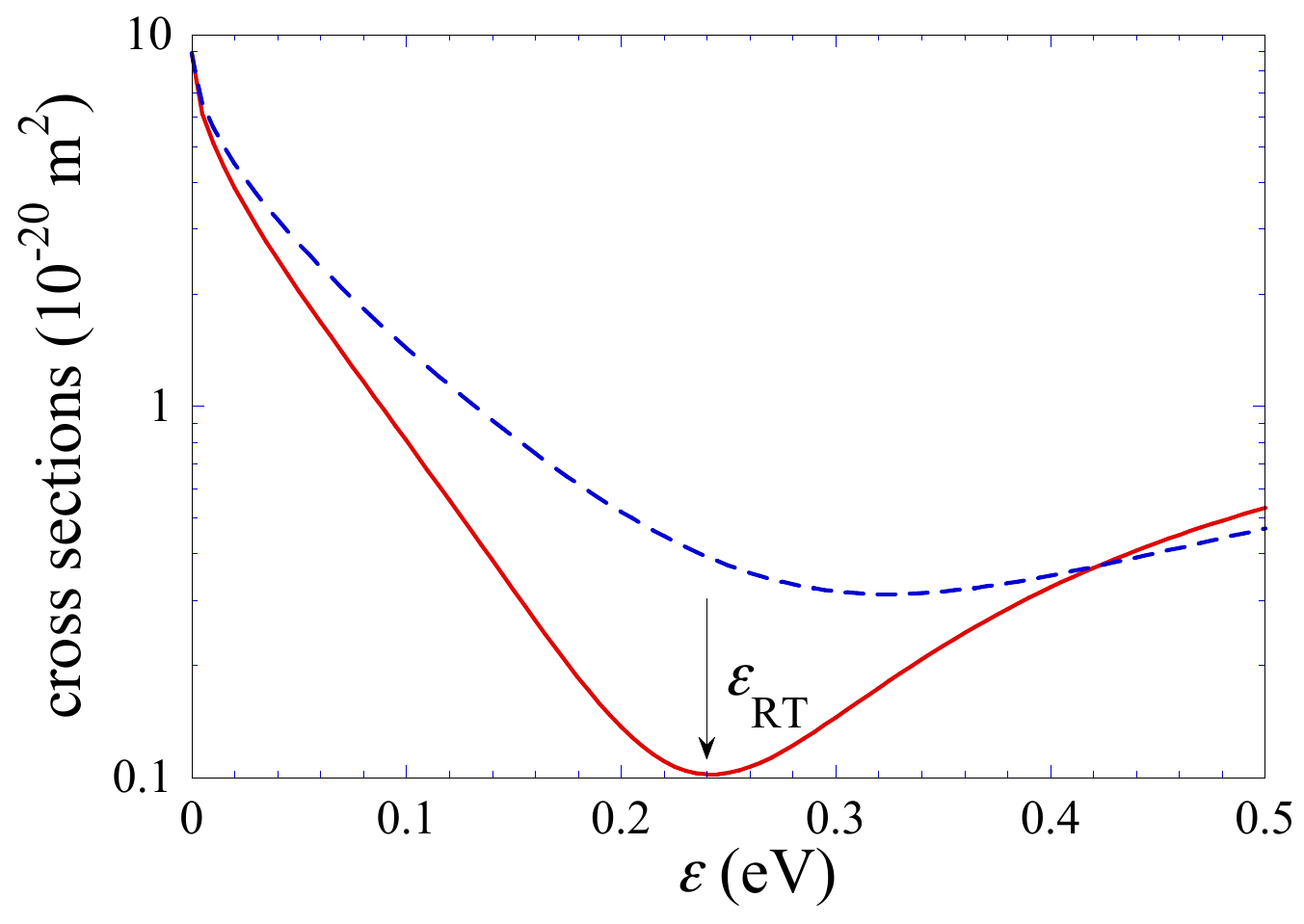}
		\caption{\small Typical electron-atom scattering cross sections~\cite{Wey88}. Solid line: momentum-transfer cross section. Dashed line: total cross section. The Ramsauer-Townsend minimum occurs at an energy \(\varepsilon_\mathrm{RT}\approx 230\,\)meV.\label{fig:figure3}}
	\end{figure}
	
	The electron-atom cross section for argon, on the contrary, is large but rapidly decreases with increasing energy~\cite{Wey88,OMa63,Suz90}, as can be seen in~\figref{fig:figure3}. Therefore, the kinetic energy shift \(E_k(N)\) produces most of the deviations from the prediction of the classical kinetic theory, but the quantum self-interference cannot be neglected at all. Moreover, as we have carried out measurements close to the critical temperature, also the correlation among scatterers must be accounted for.

	\begin{figure}[t!]
		\centering\includegraphics[width=\columnwidth]{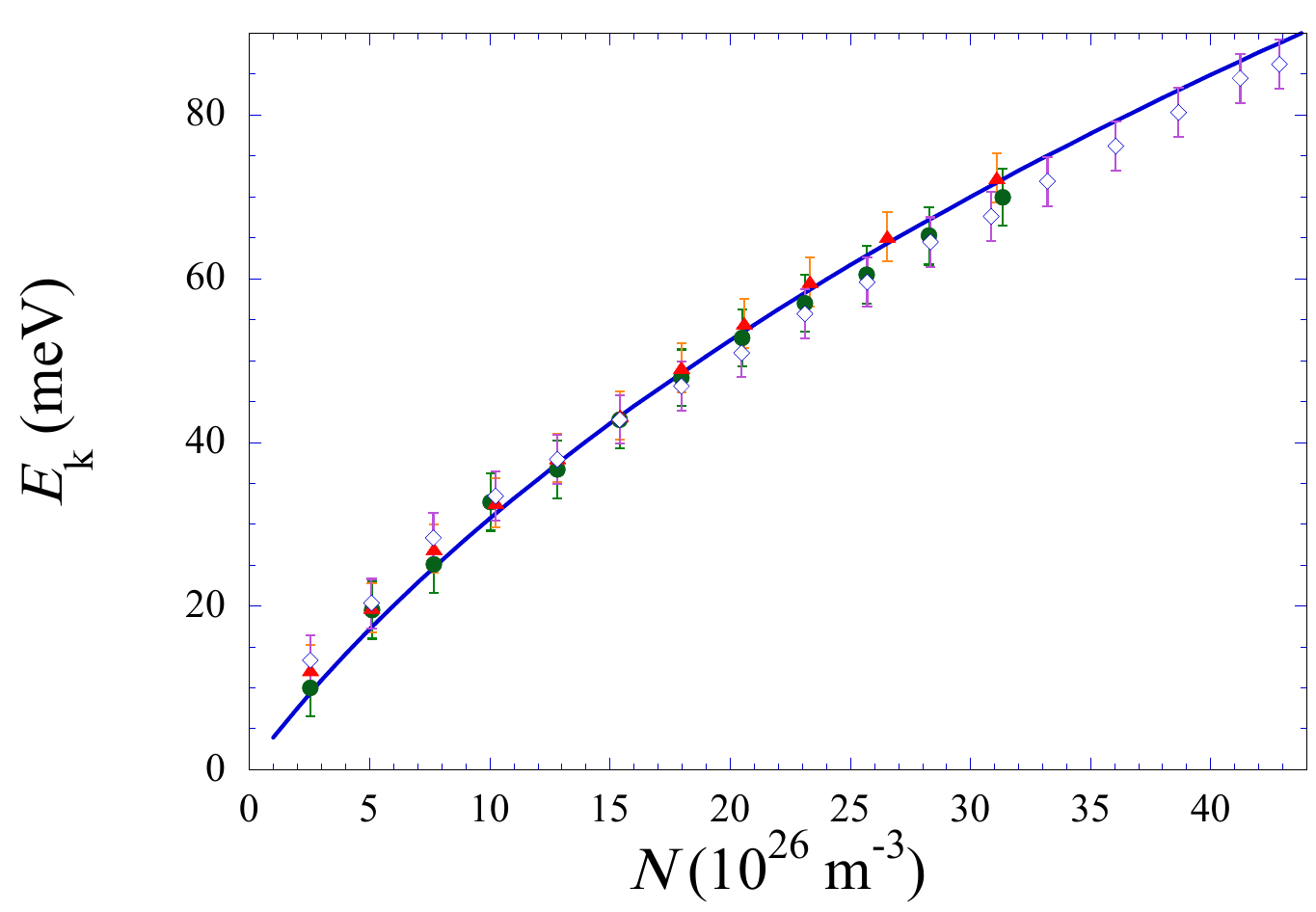}
		\caption{\small Comparison of the density dependence of the experimentally determined energy shift value \(E_k\) for several temperatures with the theoretical prediction.  The temperatures are: \(T=142.6\,\)K (triangles), \(T=177.3\,\)K (diamonds), and \(T=199.7\,\)K (circles). Solid line:  Wigner-Seitz model~\eqnref{eq:ekk0}.
			\label{fig:figure4}}
	\end{figure}
	In our first paper on electron drift in dense argon gas at \(T=162.7\,\)K~\cite{Bor92}, we determined the \(E_k(N)\) values necessary to obtain agreement between the experimentally determined \(\mu_0N\) and the theoretical prediction obtained by implementing the formulas of the heuristic model~\eqnref{eq:muN} through~\eqnref{eq:SK}, and succesfully compared it to the prediction of the WS model~\eqnref{eq:evko} and~\eqnref{eq:ekk0}. We have followed the same approach here and the results are shown in~\figref{fig:figure4}.
	
	\begin{figure}[t!]
		\centering
		\includegraphics[width=\columnwidth]{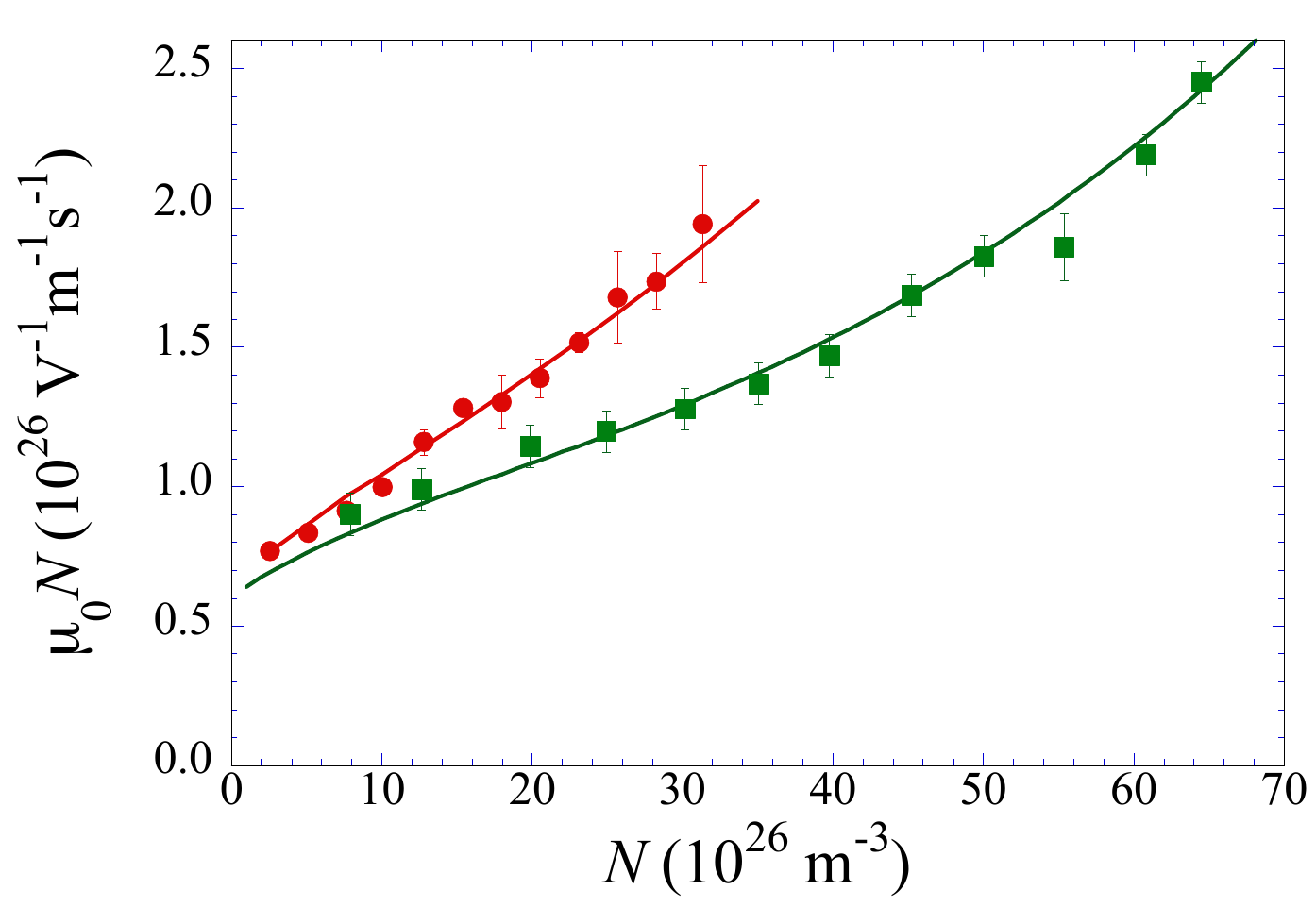}
		\caption{\small Density dependence of the zero-field density-normalized mobility \(\mu_0 N\) for \(T=199.7\,\)K (circles) and for \(T=162.7\,\)K (replotted anew from Ref.~\cite{Bor92}). Solid lines: prediction of the heuristic model using the theoretical prediction for \(E_k(N)\), shown by the  line in~\figref{fig:figure4}.
			\label{fig:figure5}}
	\end{figure}
	The agreement between the theoretical prediction for \(E_k(N)\) and the experimentally determined value is confirmed to be excellent at all investigated temperatures, even below the critical temperature. The results obtained for the isotherm \(T=152.7\,\)K, i.e., for the one closest to \(T_c,\) will be presented and discussed later because of the interesting features that emerge when the density reaches very high values.
	\begin{figure}[b!]\centering\includegraphics[width=\columnwidth]{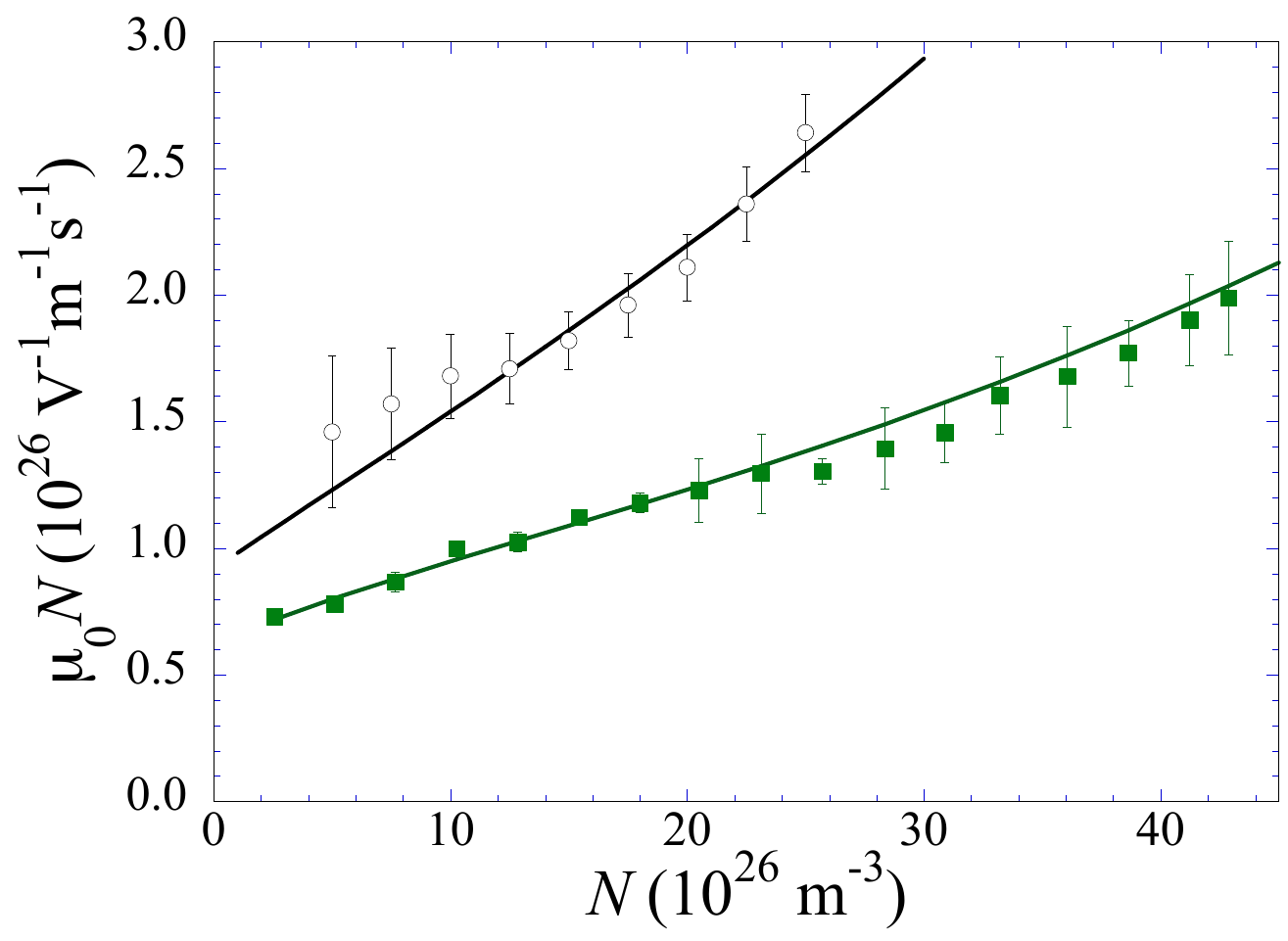}
		\caption{\small Density dependence of the zero-field density-normalized mobility \(\mu_0 N\) for \(T=300\,\)K (replotted anew from Ref.~\cite{Bartels1973}) (circles)  and for \(T=177.3\,\)K (squares). Solid lines: prediction of the heuristic model
			using the theoretical prediction for \(E_k(N)\), shown by the  line in~\figref{fig:figure4}.\label{fig:mu0NvsTT177300K}
			\label{fig:figure6}}
	\end{figure}
	\begin{figure}[t!]\centering
		\includegraphics[width=\columnwidth]{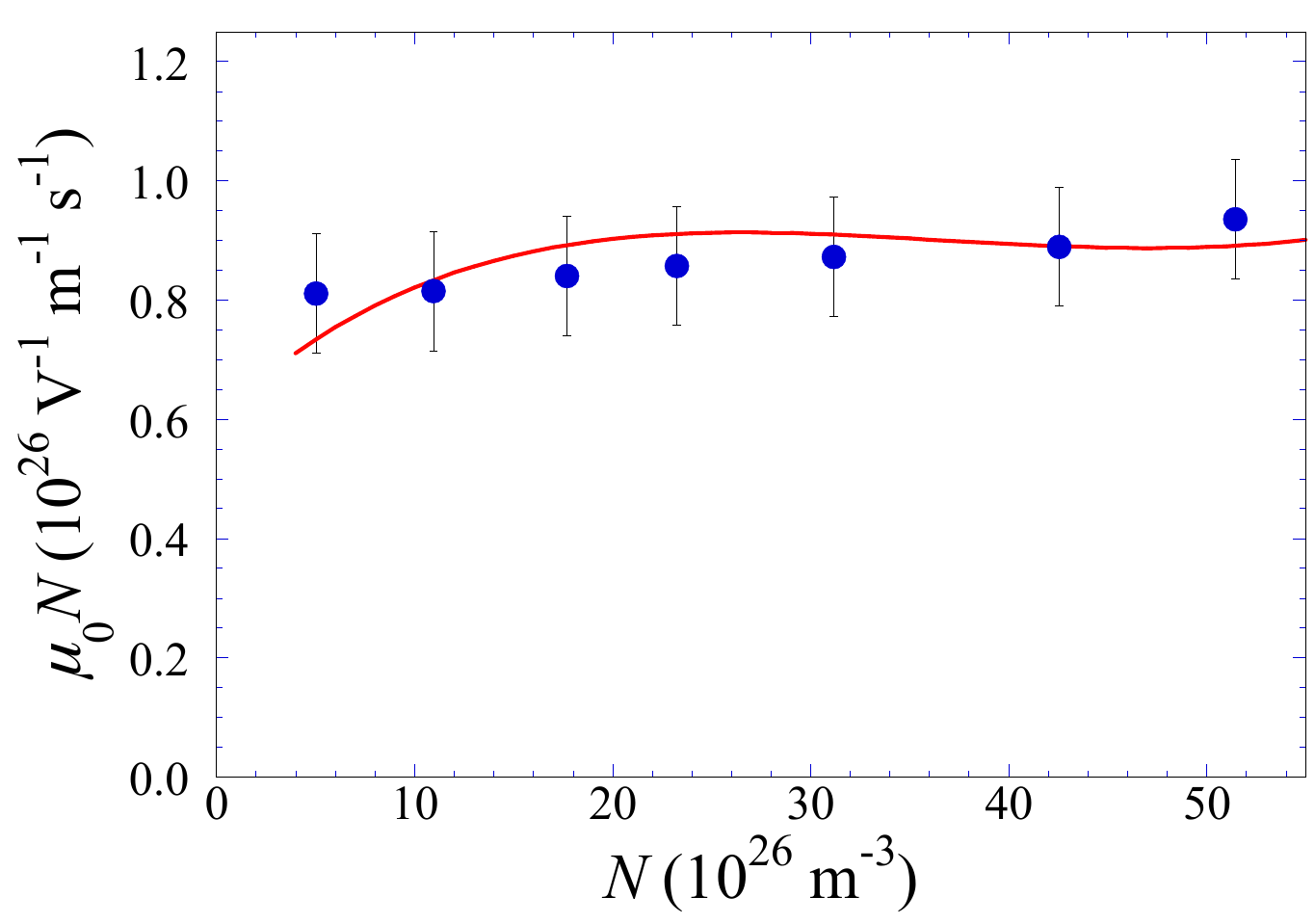}
		\caption{\small Density dependence of the zero-field density-normalized mobility \(\mu_0 N\) for \(T=152.15\,\)K. Solid lines: prediction of the heuristic model
			using the theoretical prediction for \(E_k(N)\), shown by the  line in~\figref{fig:figure4}.\label{fig:figure7}}
	\end{figure}
	
\begin{figure}[b!]\centering\includegraphics[width=\columnwidth]{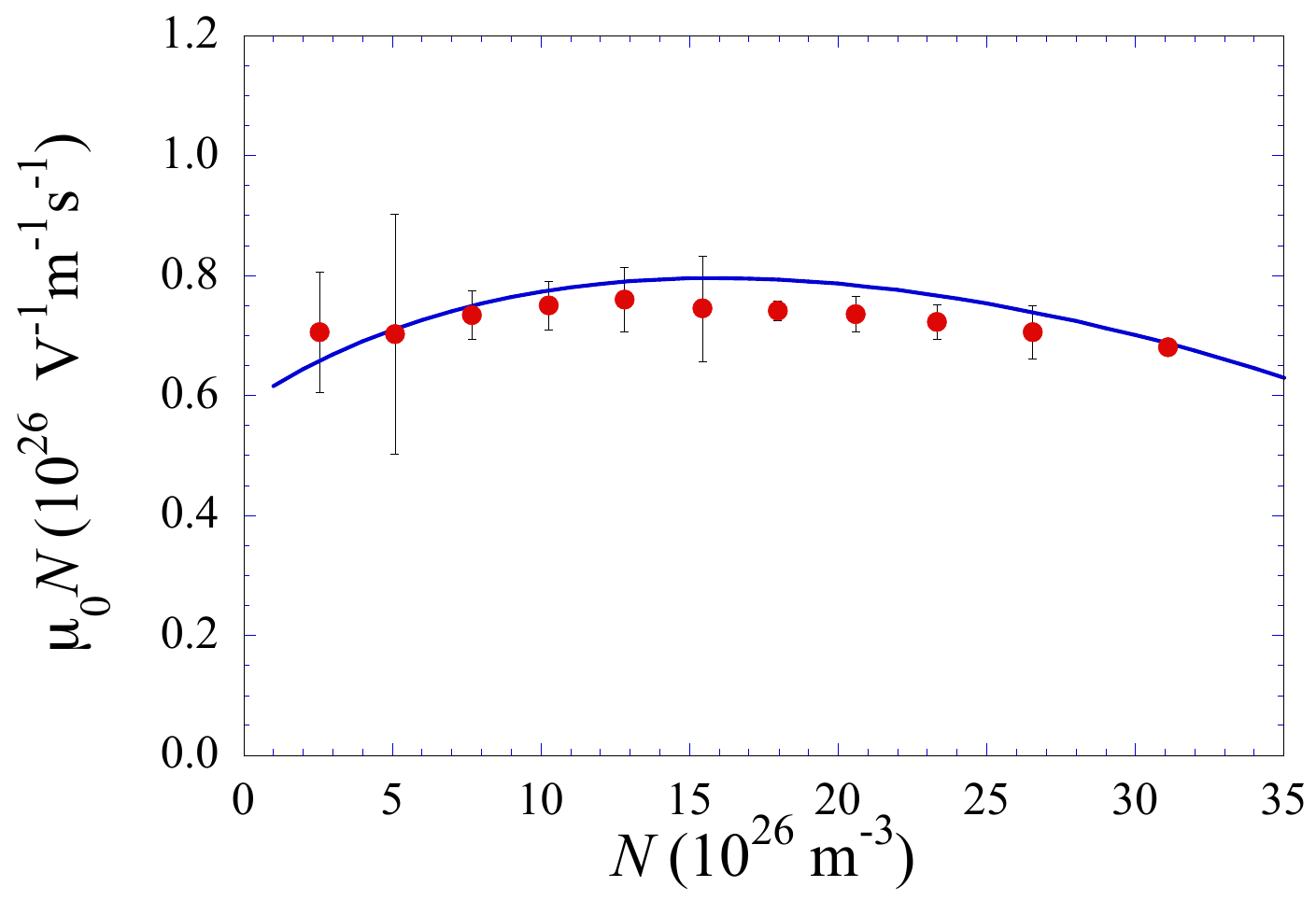}
		\caption{\small Density dependence of the zero-field density-normalized mobility \(\mu_0 N\) for \(T=142.6\,\)K. Solid lines: prediction of the heuristic model
			using the theoretical prediction for \(E_k(N)\), shown by the  line in~\figref{fig:figure4}.\label{fig:mu0NvsTT143K}
			\label{fig:figure8}}
	\end{figure}

	Due to the extremely good agreement between the theoretical and experimental \(E_k(N),\), we can now compute \(\mu_0 N\) using the theoretical prediction of the WS model~\eqnref{eq:ekk0}  and compare it with its experimental determination.
	We show in~\figref{fig:figure5} through~\figref{fig:figure8} the results of such computations.  The good agreement of the predictions of the heuristic model with the experimental \(\mu_0 N\) data is evident at all temperatures, including the old room temperature data by Bartels~\cite{Bartels1973}, shown in~\figref{fig:mu0NvsTT177300K}, and the data below \(T_c,\) plotted in~\figref{fig:figure8}.
	
	We now have to make two observations.
	The first one concerns the data for \(T=152.15\,\)K.  Close to \(T_c,\) we have been able to reach densities much higher than elsewhere. We will postpone the analysis of such high-density data to a next,  dedicated section because it will give interesting pieces of information about the transition  of the description of the  electron transport properties in a dense gas in terms of ``discrete'' collisional events to one  in terms of a ``continuum'' approach.
	
	The second remark is about the data below \(T_c.\) We note that in this case \(\mu_0N \) does not increase with \(N\) as for the higher temperatures, although it is satisfactorily reproduced by the heuristic model.  Apparently, there is a discrepancy between what was stated in the literature, i.e., argon is an {\em attractive} gas, and our data at \(T=142.6\,\)K. However, we note that measurements performed at \(T=90\,\)K, although in a very restricted density range, showed that \(\mu_0N\) is actually decreasing with \(N\)~\cite{Robertson1977}. Thus, by collecting together all pieces of information, it is evident that the value of the average derivative of \(\mu_0N \) with respect to \(N\) decreases with decreasing temperature until it eventually changes sign. We also stress the fact that this behavior is well described by the prediction of the heuristic model.
	{}
	
\begin{figure}[t!]
		\centering
		\includegraphics[width=\columnwidth]{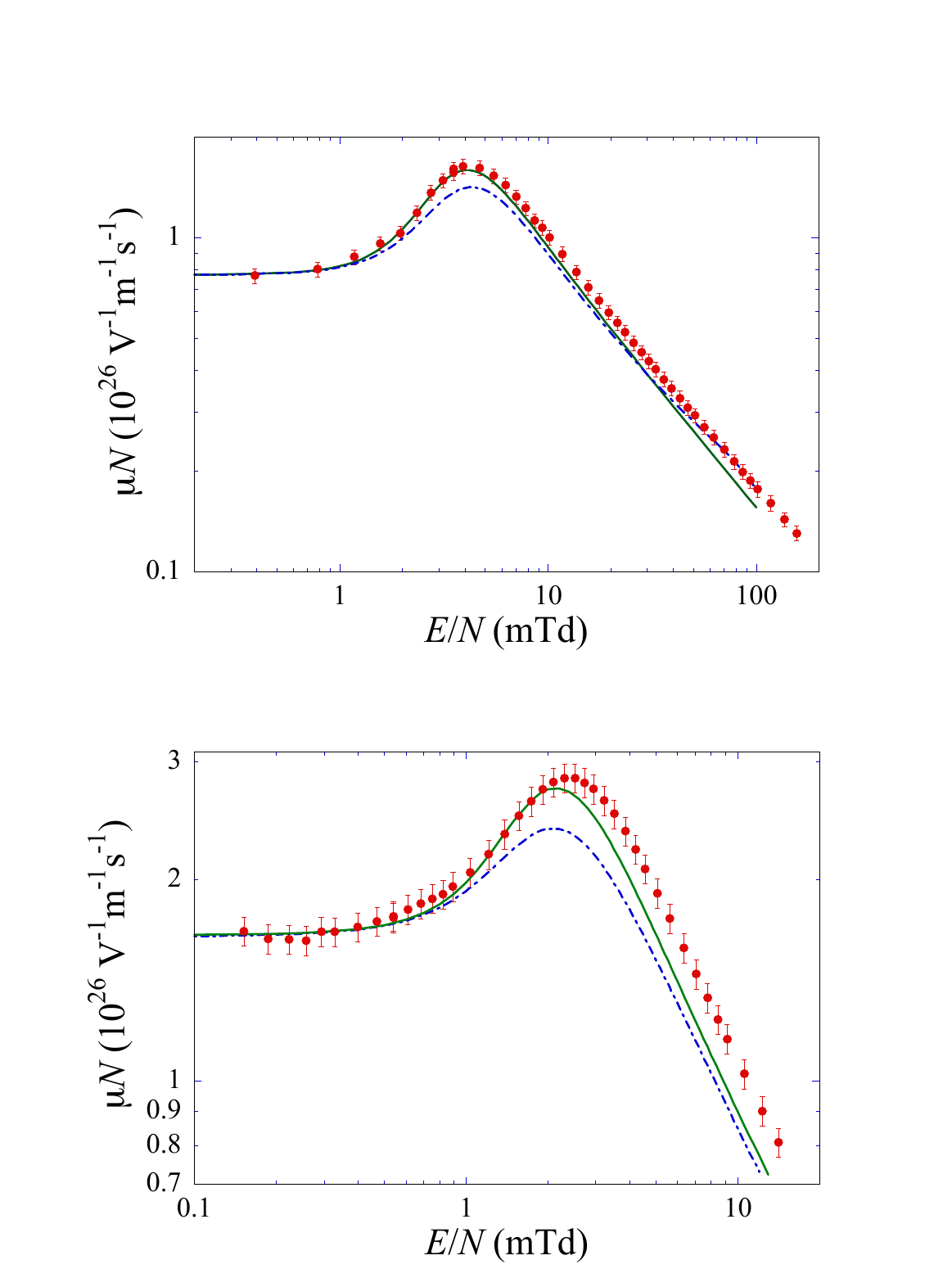} \caption{\small Density-normalized mobility \(\mu N\) vs  reduced electric field \(E/N\) at \(T=199.7\,\)K for two densities (in units of \(10^{26}\,\mbox{m}^{-3}\)): \(N=2.6\) (top panel) and \(N=28.3\) (bottom panel). The lines are the prediction of the heuristic model, using respectively using the cross sections by Suzuki~\cite{Suz90} (solid lines) and by Weyhreter {\em et al.}~\cite{Wey88} (dash-dotted lines). (\(1\,\mbox{mTd} = 10^{-24}\mbox{Vm}^2\)).
			\label{fig:figure9}}
	\end{figure}

	To conclude this section, we would like to show how the heuristic model is successful in reproducing quite accurately also the electric field dependence of the density reduced mobility \(\mu N.\) In~\figref{fig:figure9} we show the electric field dependence of \(\mu N \) at \(T=199.7\,\)K for  two values of density: one low and one high. Although the kinetic energy shift \(E_k(N)\) is determined by seeking agreement with the data at weak electric fields, the heuristic model nevertheless reproduces the electric field dependence of mobility with good accuracy. However, the overall agreement depends on the choice of the cross section. For example, the cross section reported by Suzuki {\em et al.}~\cite{Suz90} has been determined by analyzing swarm data, whereas the cross section by Weyhreter {\em et al.}\citep{Wey88} has been measured in a crossed-beam experiment. As a consequence, both the energy range of validity and the energy resolution of these cross sections differ from each other.
	
	\subsection{Heating electrons by increasing the density at constant temperature and electric field}\label{sec:heatingNvsT}
	In this section, we will prove that the average electron energy in the dense gas can be changed not only by varying  the gas temperature or the electric field acting upon the electrons, as can be intuitively expected, but also by varying the density due to the contribution of the density-dependent kinetic energy shift \(E_k(N)\).

	\subsubsection{Analysis of the zero-field density-normalized mobility}\label{sec:mu0nanalysis}
	\begin{figure}[b!]
		\centering\includegraphics[width=\columnwidth]{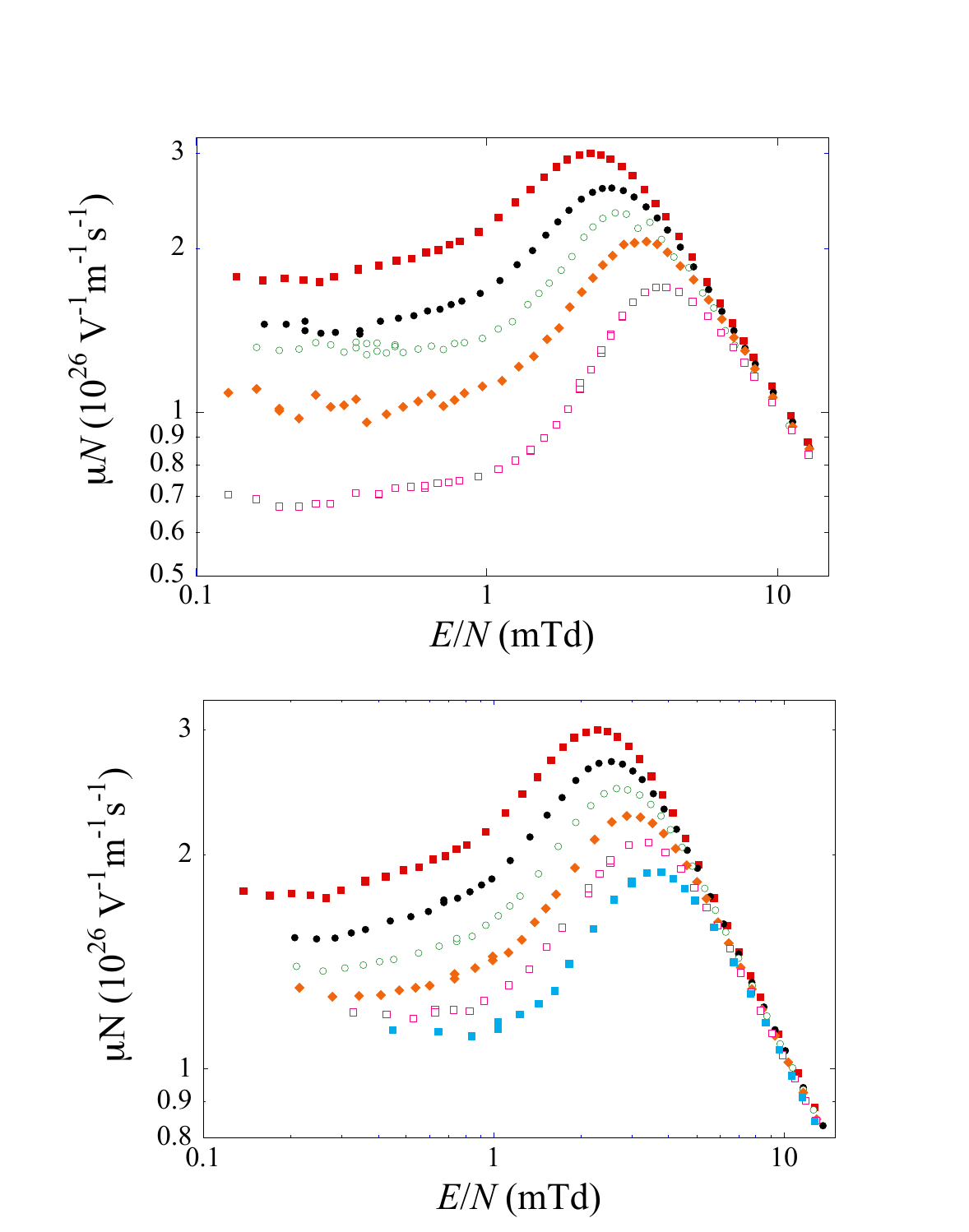}
		\caption{\small Reduced electric field dependence of the density-normalized mobility \(\mu N\) for several  \(T\) and \(N.\) Top panel: \(N\approx 31\times 10^{26}\,\mbox{m}^{-3}\) at \(T\,(\mbox{K}) = 199.7,\,177.3,\,162.7,\,152.7,\, 142.6\) (from top). Bottom panel: \(T=199.7\,\)K and \(N\left(10^{26}\,\mbox{m}^{-3}\right)= 31.3,\,25.7,\,20.5,\, 15.4,\,10.1,\,5.12\) (from top). 
			\label{fig:figure10}}
	\end{figure}
	A first hint at understanding how electrons can be heated by increasing the density at constant field and temperature is obtained by rationalizing the density dependence of \(\mu_0 N\) at constant temperature. Actually, according to ~\eqnref{eq:mu0Ncl}, \(\mu_0N \) can be considered as a kind of thermal average of the inverse cross section. Roughly speaking, its thermal average can be approximated by the cross section value at the average electron energy  \( \langle\sigma_\mathrm{mt}^{-1}\rangle\simeq \sigma_\mathrm{mt}^{-1}(\langle \varepsilon\rangle).\)
	
	Taking into account the specific energy dependence of \(\sigma_\mathrm{mt}\) shown in~\figref{fig:figure3}, it is clear that the cross section has to be evaluated at an average energy that increases with increasing density in order to reproduce the observed density dependence of \(\mu_0N.\)
	
	A striking visual confirmation of the similar effectiveness of temperature and density to heat electrons up  (or, more precisely, to increase their average energy) is obtained by inspecting~\figref{fig:figure10}. 
	
	On one hand, the top panel displays the density-normalized mobility \(\mu N\) as a function of the reduced electric field \(E/N\) at constant  \(N\approx 31\times 10^{26}\,\)m\(^{-3}\) for several temperatures. The succession of the curves is naturally ordered according to \(T:\) the higher is the temperature, the higher is the corresponding \(\mu N.\) Once again, the rationalization of this behavior is simple, namely, the increase of the temperature brings about an increase of the average electron energy to which a decrease of the average cross section value corresponds, thereby yielding an increase of the mobility. 
	
	On the other hand, the bottom panel shows a similar succession of \(\mu N\) curves obtained at constant \(T=199.7\,\)K for different densities. In this case, the curves ordering follows the density ordering: the larger is the density, the higher is the mobility. Increasing \(N\) adds an increasingly larger \(E_k(N)\) contribution to the average electron energy and, once more, the value of the average cross section decreases while the mobility increases. 
	
\begin{figure}[b!]
		\centering
		\includegraphics[width=\columnwidth]{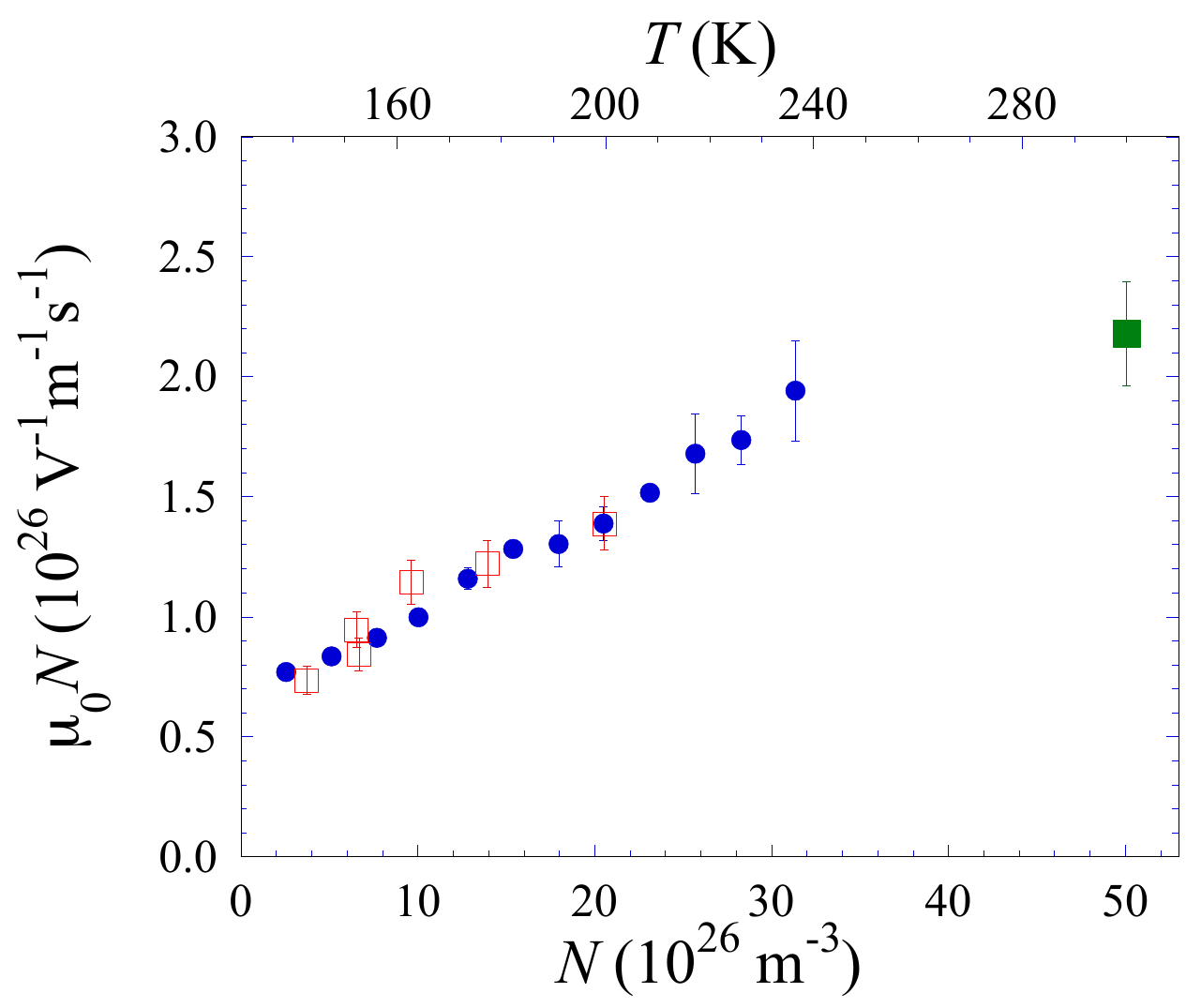}
		\caption{\small 
			Increase of the zero-field density normalized mobility \(\mu_0 N\) with increasing \(N\) at constant \(T=199.7\,\)K (circles, bottom scale) or with increasing \(T\)  at constant \(N\approx 20\times 10^{26}\,\)m\(^{-3}\) (squares, top scale). Solid square: plotted anew from Ref.~\cite{Bartels1973}. \label{fig:DensityHeatingEffectonmu0N}}
	\end{figure}

\begin{figure}[t!]
		\centering
		\includegraphics[width=\columnwidth]{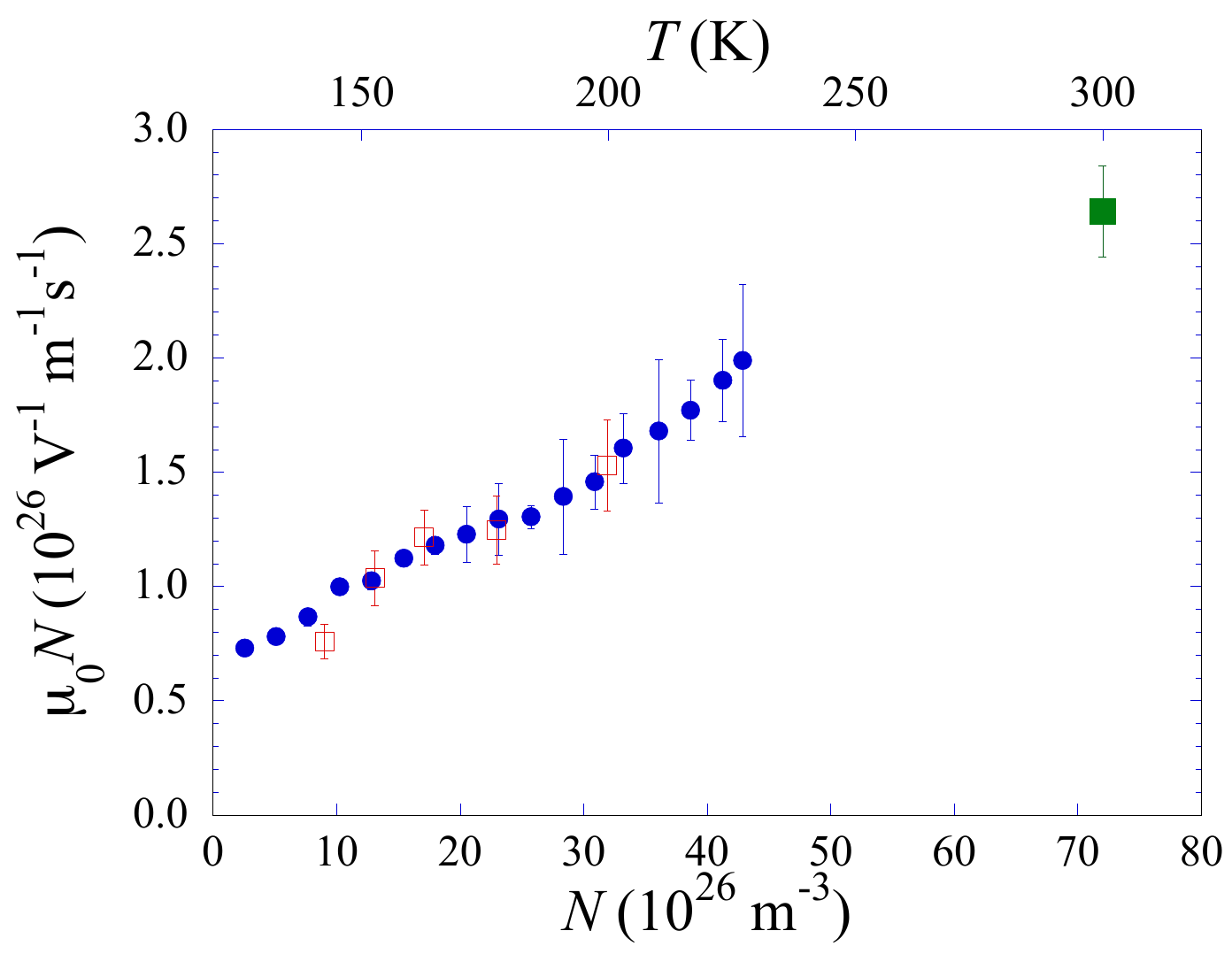}
		\caption{\small  
			Increase of the zero-field density normalized mobility \(\mu_0 N\) with increasing \(N\) at constant \(T=177.3\,\)K (circles, bottom scale) or with increasing \(T\)  at constant \(N\approx 25\times 10^{26}\,\)m\(^{-3}\) (squares, top scale). Solid square: plotted anew from Ref.~\cite{Bartels1973}.
			\label{fig:figure12}}
	\end{figure}

	This behavior is followed at all temperatures and densities~\cite{Bor03} and is summarized  in~\figref{fig:DensityHeatingEffectonmu0N} and in~\figref{fig:figure12}.
	In these figures we show  simultaneously the zero-field density normalized mobility \(\mu_0 N\) measured at \(T=199.7\,\)K  as a function of \(N\) and the value of \(\mu_0 N\) for \(N\approx 20\times 10^{26}\,\)m\(^{-3}\) as a function of \(T\) and  at \(T=177.3\,\)K for \(N\approx 25\times 10^{26}\,\)m\(^{-3}\) By plotting these data in the same picture by using two different abscissas we make evident that both temperature and density act so as to drive \(\mu _0N\) in a similar way. 
	

	\subsubsection{Analysis of the reduced field at the mobility maximum}\label{sec:EsuNmaxAnalysis}
\begin{figure}[t!]
		\centering
		\includegraphics[width=\columnwidth]{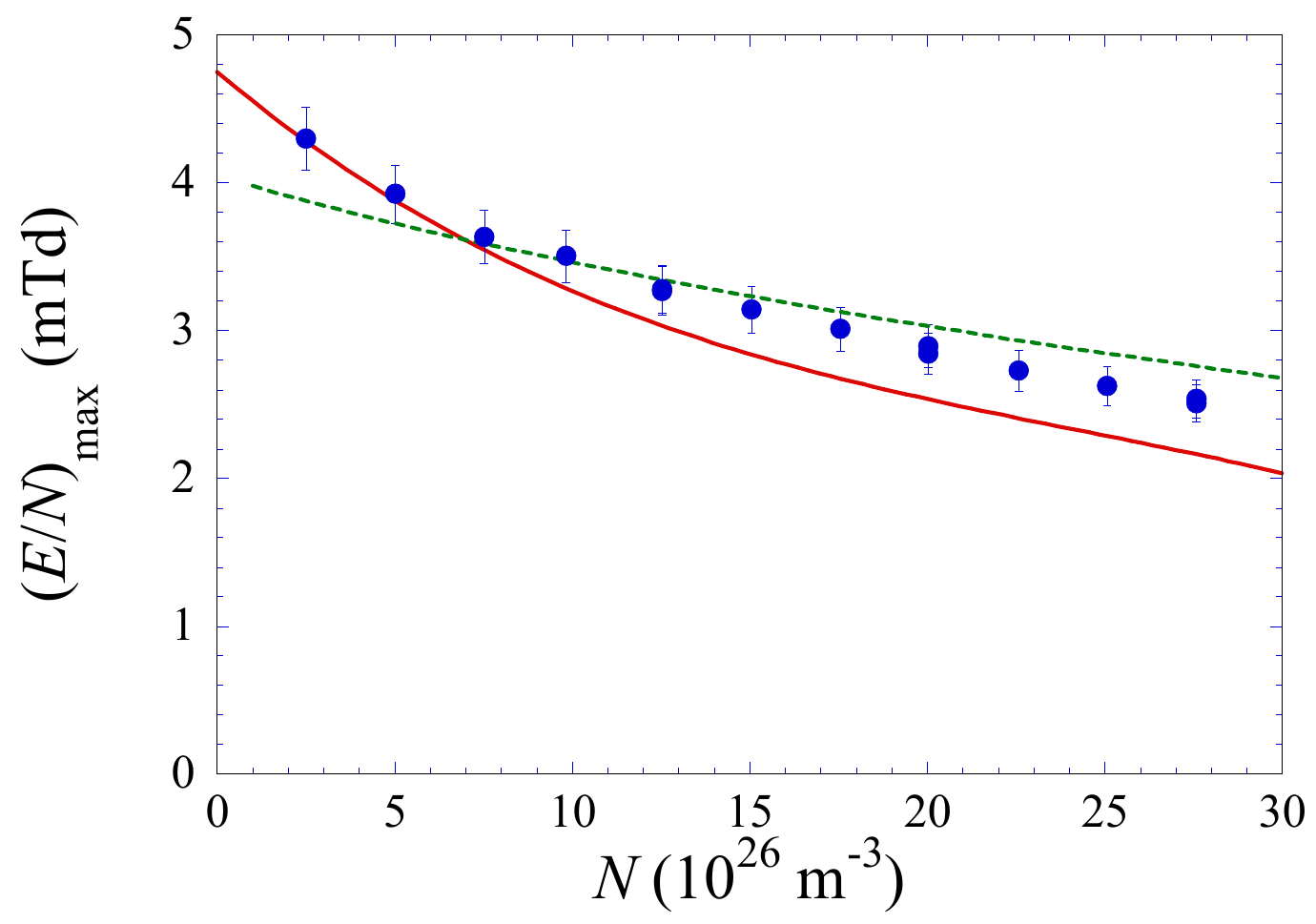}
		\caption{ \small Density dependence of the value of the reduced field \((E/N)_\mathrm{max}\) at the mobility maximum for \(T=199.7\,\)K. Solid line: prediction of the heuristic model using the theoretical value for the kinetic energy shift \(E_k(N),\)~\eqnref{eq:ekk0}, and the  cross section of Weyhreter {\em et al.}~\cite{Wey88}. Dashed line: approximate prediction,~\eqnref{eq:enmax} with \(\sigma_0=2.3\times 10^{-20}\,\mbox{m}^2\).\label{fig:figure13}}
	\end{figure}

	The maximum shown by \(\mu N\)  at the reduced electric field strength \((E/N)_\mathrm{max}\) is a strong evidence that, for this field value, the average electron energy \(\langle \varepsilon\rangle\) equals the energy \(\varepsilon_\mathrm{RT}\) of the Ramsauer-Townsed minimum in \(\sigma_\mathrm{mt}\). 
	In~\figref{fig:figure13} we show the density dependence of \((E/N)_\mathrm{max}\) for \(T=199.7\,\)K. We note that \((E/N)_\mathrm{max}\) decreases with increasing \(N.\)
	As the average energy gained by the electrons from the field over one mean free path is \(eE\ell\propto e(E/N\sigma_\mathrm{mt})\), and it decreases with increasing the density,  there must be another density-dependent source of electron energy to keep the average energy constant at the value of the Ramsauer-Townsend minimum. This source is evidently the density-dependent kinetic energy shift \(E_k(N).\)
	
	We can prove it in a simplified way by observing that, 
	for \(E/N\approx \left(E/N\right)_\mathrm{max},\) the average electron energy \(\langle \varepsilon\rangle\) is significantly higher than the thermal contribution.
	In this case the distribution function reduces to
	\begin{equation}g(\varepsilon) \approx
		A\exp{\left\{-\int\limits_0^\varepsilon
			\frac{6 m z}{M(eE/N\sigma^\star_\mathrm{mt})^2}\,\mathrm{d}z\right\}}
		\label{eq:davred}
	\end{equation}
	To further simplify the analysis, we assume that, for \(\varepsilon\) close to \(\varepsilon_\mathrm{RT},\) \(\sigma^\star_\mathrm{mt}\) can be approximated by some average value \(\sigma_0.\)
	In this case  the integration in~\eqnref{eq:davred} can easily be performed in a closed form, yielding
	\begin{equation}
		g(\varepsilon) \approx A\exp{\left\{-C\varepsilon^2
			\right\}} 
		\label{eq:davint}
	\end{equation}
	in which
	\(C={3m}/{M\left(eE/N\sigma_0\right)^2}\) and the normalization constant is
\(		A=
{2C^{3/4}}/{\Gamma(3/4)}\)
	where \(\Gamma\) is the Euler's gamma function.
	
	The average electron energy is then given by  
	\begin{eqnarray}
		\langle\varepsilon
		\rangle&=&  A\int\limits_0^\infty \varepsilon^{3/2} \exp{\left\{-C\varepsilon^2\right\}}\,\mathrm{d}\varepsilon\nonumber\\
		&=&\frac{\Gamma(5/4)}{\Gamma(3/4)C^{1/2}}\approx 115.3\left(\frac{eE}{N\sigma_0}\right)
		\label{eq:eaveAtRT}
	\end{eqnarray}
	We note that this contribution is actually the average energy acquired by the electron from the field over one mean free path.
	To this term we now add the thermal contribution \((3/2)k_\mathrm{B}T\) and the density-dependent kinetic energy shift \(E_k(N).\)
	At the energy \(\varepsilon_\mathrm{RT}
	\), i.e., 
	for
	\(\left(E/N\right)_\mathrm{max},\) we thus have
	\begin{equation}
		\varepsilon_\mathrm{RT}=\frac{115.3}{\sigma_0}
		\left(\frac{E}{N}\right)_\mathrm{max}
		+\frac{3}{2}\frac{k_\mathrm{B}}{e}T+E_k(N)
		\label{eq:eRTenmax}
	\end{equation}
	if energies are expressed in eV units.
	\eqnref{eq:eRTenmax} is inverted to yield
	\begin{equation}
		\left(\frac{E}{N}\right)_\mathrm{max}=\frac{\sigma_0}{115.3}\left[\varepsilon_\mathrm{RT}-\frac{3}{2}\frac{k_\mathrm{B}}{e}T -E_k(N)
		\right]
		\label{eq:enmax}
	\end{equation}
	\eqnref{eq:enmax} is shown in~\figref{fig:figure13} as the dashed line if we use a reasonable value for \(\sigma_0=2.3\times 10^{-20}\,\)m\(^2.\) The agreement with the experimental data is very satisfactory in spite of the approximations  we have adopted.
	
	In~\figref{fig:figure13} we also plot the prediction of the full heuristic model as the solid line. In this case, we used the theoretical value of \(E_k(N)\),~\eqnref{eq:ekk0}, and the cross section by Weyhreter {\em et al.}~\cite{Wey88}. Even in this case the agreement is excellent. It is, however, worth noting that discrepancies among the different cross sections found in literature~\cite{OMa63,Wey88,Suz90} may yield significant differences in the predicted values of \((E/N)_\mathrm{max},\) as can be easily deduced by inspecting~\figref{fig:figure9}, although the conclusions about the physics of the problem are not spoiled.

	\begin{figure}[b!]
		\centering		\includegraphics[width=\columnwidth]{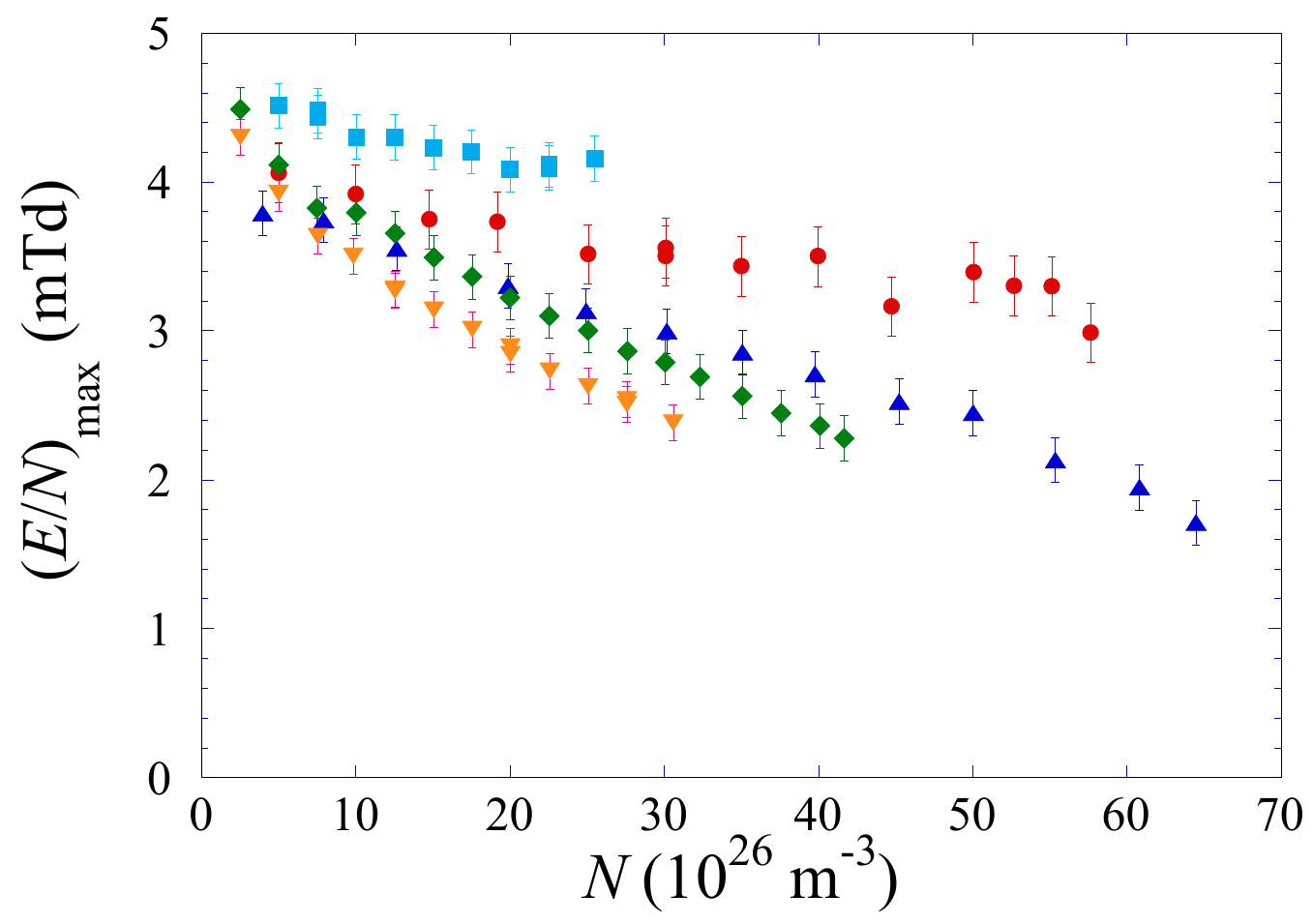} \caption{\small Density dependence of the reduced electric field at the mobility maximum \((E/N)_\mathrm{max}\) for several temperatures. \(T\,(\mbox{K}) =142.6,\, 152.7,\,162.7,\,177.3,\,199.7\) (from top). The data at \(T=162.7\,\)K are plotted anew from Ref.~\cite{Bor92}.\label{fig:figure14}}
	\end{figure}
	An additional consequence of the previous analysis is that we have to expect that, for a given density, \((E/N)_\mathrm{max}\) must decrease with increasing temperature because of the increasingly larger contribution of the thermal energy. This is actually the case, as shown in~\figref{fig:figure14}.  The decrease of \((E/N)_\mathrm{max}\) with increasing \(T\) for any density is very well confirmed. Moreover, the agreement between experiment and theory for each temperature is comparable to that shown in~\figref{fig:figure13}.  
	
	As anticipated for the analysis of \(\mu_0 N\), a detailed analysis of the \((E/N)_\mathrm{max} \) for the case of the \(T=152.7\,\)K, the temperature closest to the critical one, will be postponed to a next section.
	
	\begin{figure}[t!]
		\centering
		\includegraphics[width=\columnwidth]{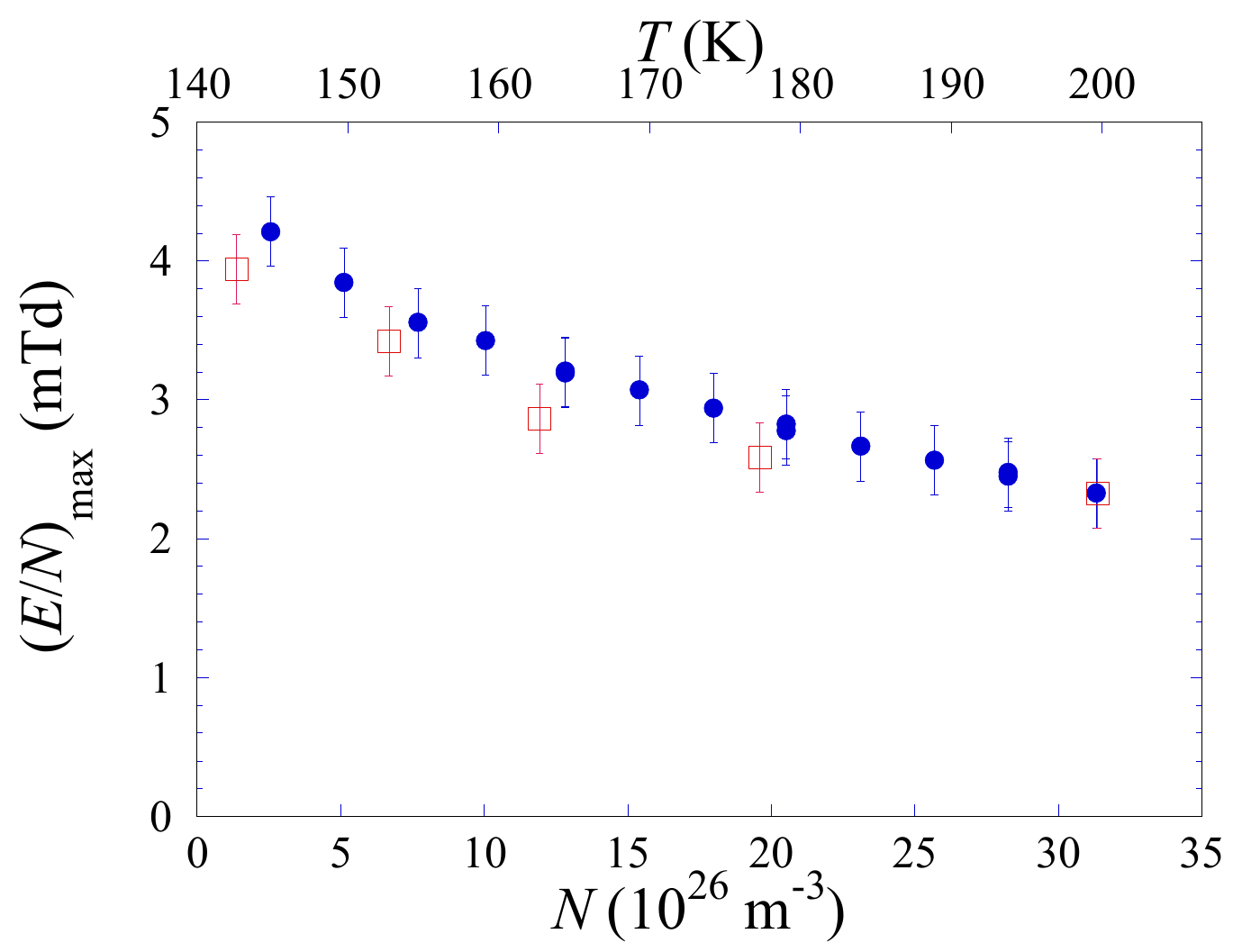}\caption{\small Decrease of the reduced electric field at the mobility maximum \((E/N)_\mathrm{max}\)
			with increasing \(N\) at constant  \(T=199.7\,\)K (circles, bottom scale) or 
			with increasing \(T\) at constant  \(N\approx 31\times 10^{26}\,\)m\(^{-3}\)(squares, top scale). 
			\label{fig:figure15}}
	\end{figure}
	As we have previously done by dealing with \(\mu_0N\), we can now prove the similar effect of temperature and density as far as the increase of the average electron energy is concerned by considering the behavior of \((E/N)_\mathrm{max}\). 
	\begin{figure}[b!]
		\centering
		\includegraphics[width=\columnwidth]{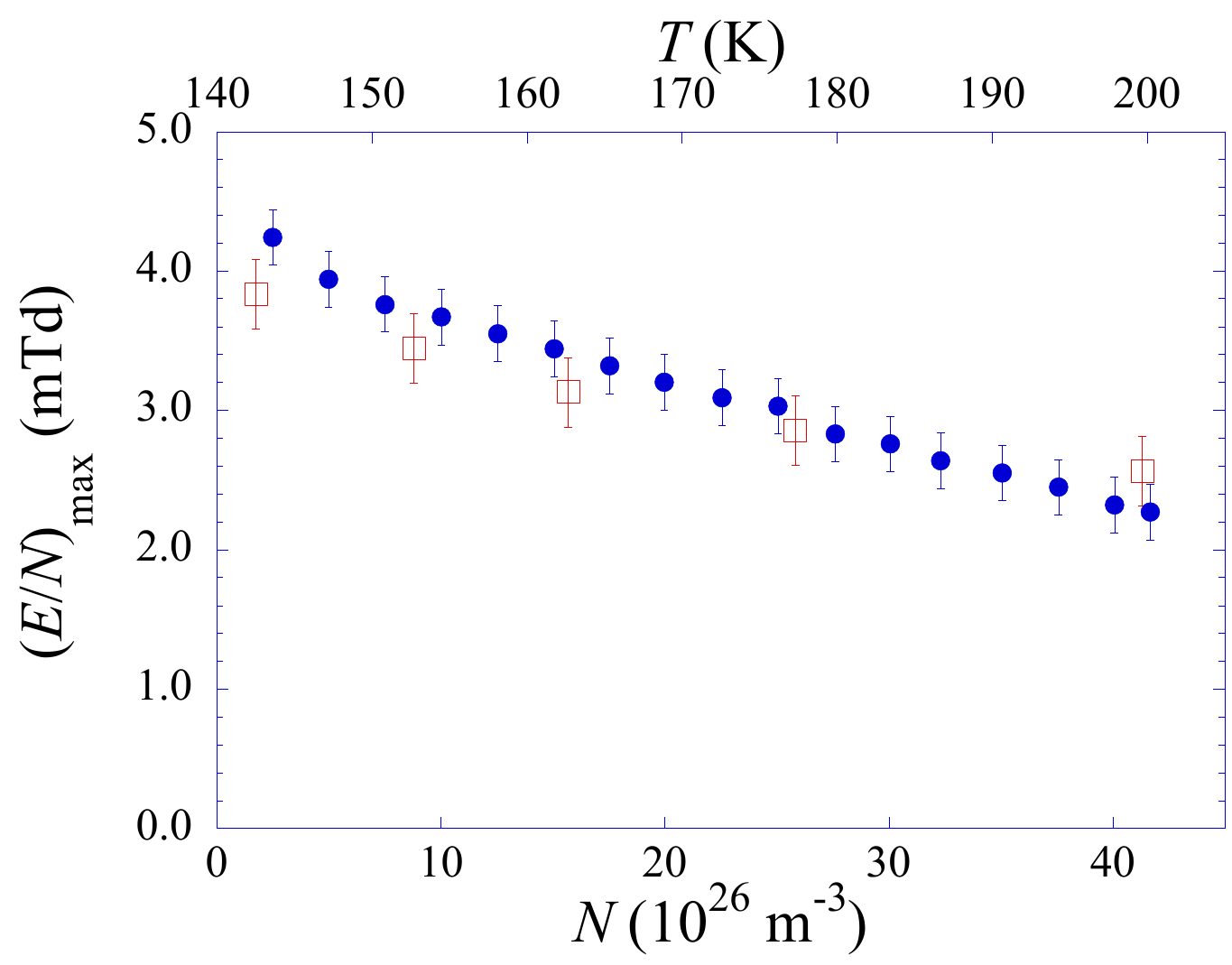}\caption{\small Decrease of the reduced electric field at the mobility maximum \((E/N)_\mathrm{max}\)
			with increasing \(N\) at constant  \(T=177.3\,\)K (circles, bottom scale) or 
			with increasing \(T\) at constant  \(N\approx 25\times 10^{26}\,\)m\(^{-3}\)(squares, top scale). 
			\label{fig:figure16}}
	\end{figure}
	In~\figref{fig:figure15} we simultaneously summarize 
	the density dependence of \((E/N)_\mathrm{max}\) at the constant \(T=199.7\,\)K, and the temperature dependence of \((E/N)_\mathrm{max}\) at the constant \(N\approx 31\times 10^{26}\,\)m\(^{-3}\). Simlarly, in~\figref{fig:figure16} we plot the density dependence of \((E/N)_\mathrm{max}\) for \(T=177.3\,\)K and its temperature dependence for \(N\approx 25\times 10^{26}\,\)m\(^{-3}.\)
	The electric field at the mobility maximum decreases with density if the temperature is kept constant. At the same time, it also decreases with temperature if the density is maintained constant. 
	In both cases, the average electron energy is affected in the same way by temperature and density, even though the physical causes are different.

	\subsubsection{The special case of the temperature closest to the critical one: \(T=152.7\,\)K}
\begin{figure}[t!]\centering
		\includegraphics[width=\columnwidth]{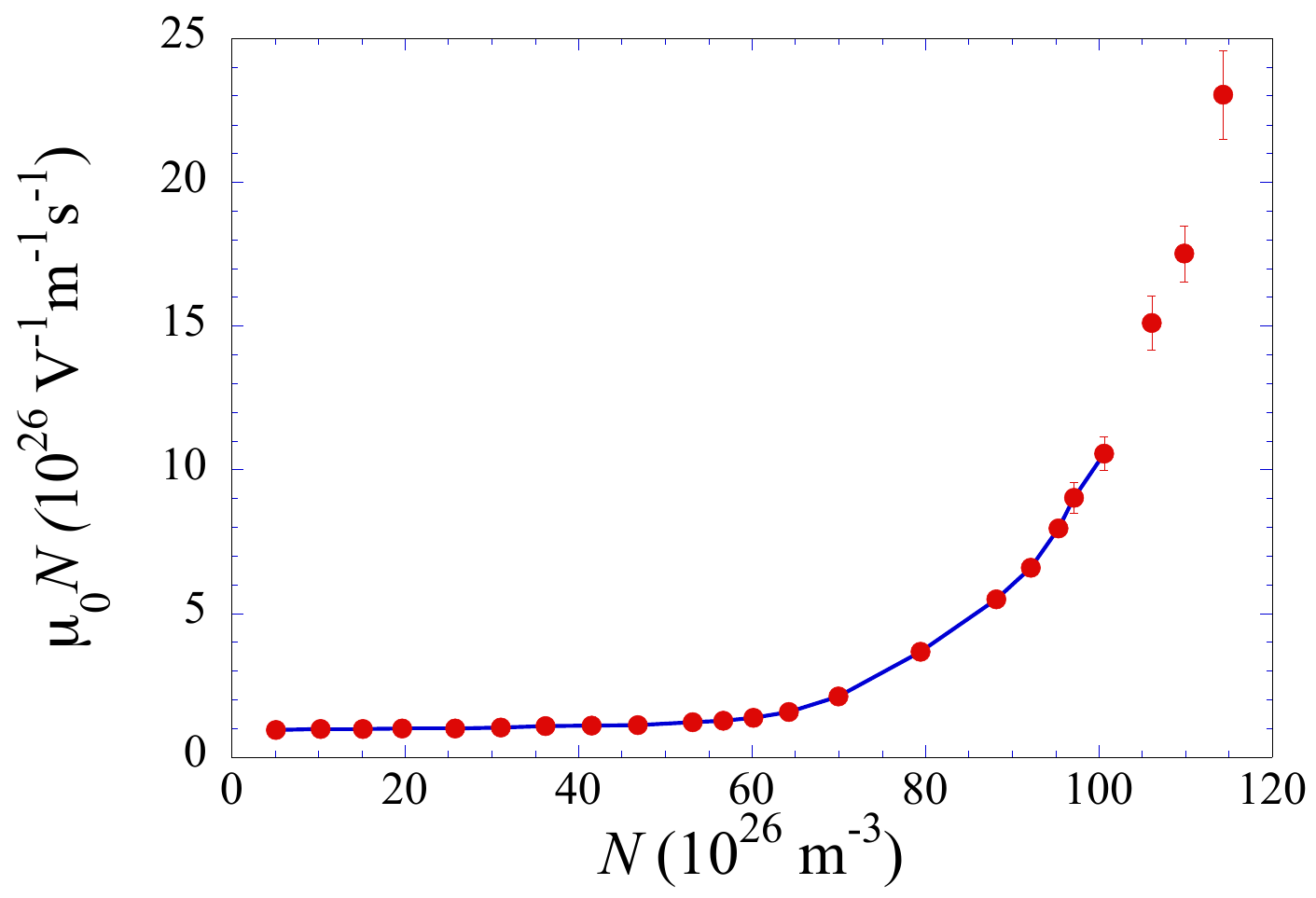}
		\caption{\small 
			Density dependence of the zero-field density normalized mobility \(\mu_0 N\) for \(T=152.7\,\)K. Solid line: prediction of the heuristic model with the kinetic energy shift \(E_K(N)\) reported in~\figref{fig:figure18}.\label{fig:figure17}}
	\end{figure}

	We deal here with the case of the \(T=152.7\,\)K isotherm, the one  closest to \(T_c\), because very high densities are reached.
	In~\figref{fig:figure17} we show the values of \(\mu_0 N\) as a function of the density. They agree with previous data at \(T=152.15\,\)K~\cite{Bor2001}. 
	
\begin{figure}[t!]\centering\includegraphics[width=\columnwidth]{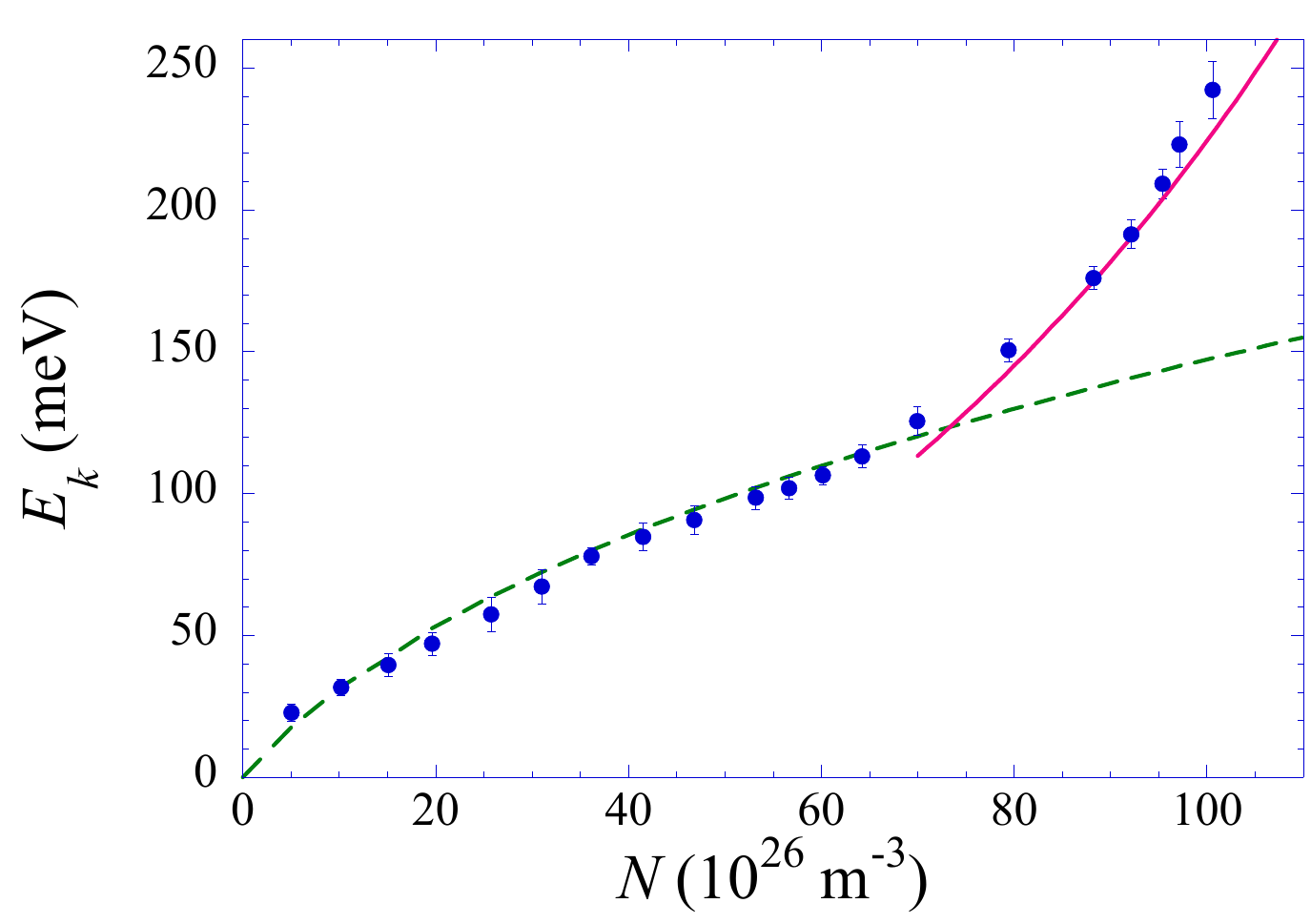}
		\caption{\small Density dependence of the kinetic energy shift \(E_k(N)\) obtained for \(T=152.7\,\)K. The points are the values to insert in the formulas of the heuristic model to reproduce the measured \(\mu_0 N\) values. The dashed line is the prediction of the WS model.  
			The solid line is obtained by subtracting the potential energy shift \(U_P(N),\) computed according to~\eqnref{eq:UP}, from the experimentally measured total energy shift \(V_0(N)\)~\cite{Rei83}. 
			\label{fig:figure18}}
	\end{figure}

	If we were using the theoretical prediction, \eqnref{eq:ekk0}, for the kinetic energy shift \(E_k(N)\) as we have used to plot the solid lines in~\figref{fig:figure5} through~\figref{fig:figure8}, the agreement of the experimental \(\mu_0 N\) with the prediction of the heuristic model would not extend beyond \(N^\star =N\approx 65\times 10^{26}\,\)m\(^{-3}.\) For higher densities \(\mu_0 N\) increases with \(N\) at a higher rate. Therefore, we have adopted the inverse, original procedure to force the agreement of the  heuristic model prediction with the experiment by seeking for the required \(E_k(N)\) for any density. The results of this kind of determination of the optimum \(E_k\) are shown in~\figref{fig:figure18}. By inserting in the equations of the heuristic model the values of \(E_k(N)\) represented by circles in this figure, the solid line in~\figref{fig:figure17} is obtained. By so doing, we have been able to push the agreement and theory up to \(N=N_M\approx 105\times 10^{26}\,\)m\(^{-3},\) well beyond the critical density \(N_c=80.8\times 10^{26}\,\)m\(^{-3}.\)
	
	By inspecting~\figref{fig:figure18} we note that for \(N\lesssim N^\star\) the experimentally determined \(E_k(N)\) values agree very well with the prediction of the WS model, here displayed by the dashed line. For \(N>N^\star\), the prediction of the WS model strongly deviates from the experimental values. Interestingly enough, \(R_\mathrm{WS}(N^\star)\approx 3.4\times 10^{-10}\,\)m, which is the hard-sphere radius of the argon-argon interaction potential~\cite{maitland}.
	For \(N>N^\star\) it appears that the experimental \(E_K(N)\) data are well represented by the solid curve, whose meaning will soon be explained 
	
	We believe that \(N^\star\) is a threshold density, beyond which the superposition of the polarization potential tails cannot be neglected. In particular, it is no longer valid the approach of computing \(E_k(N)\) according to~\eqnref{eq:evko} and~\eqnref{eq:ekk0}. In fact, \eqnref{eq:evko} is obtained by replacing the actual potential with a Hartree-Fock hard-core pseudopotential in the region where the potential is negligible between nearby atoms. This approximation is valid whenever
	the density is relatively low and the atoms quite well separated~\cite{Spri68,Her91}.

	When the density is large enough to take into account the overlap of the polarization potentials and to consider the fluid as a continuum, the polarization contribution \(U_P(N)\) to the total electron energy shift \(V_0\) in~\eqnref{eq:v0} can be written as~\cite{Her91}
	\begin{equation}
		U_P (N) \approx -\frac{\alpha eN}{2\epsilon_0}\int \limits_{R_\mathrm{WS}}^\infty  \frac {g(r)S(r)}{r^2}\,\mathrm{d}r
		\label{eq:hernUP}
	\end{equation}
	in which \(\alpha\) is the atomic polarizability of argon,  \(\epsilon_0\) is the vacuum permittivity, and \(U_P\) is expressed in eV units.   \(g(r)\)
	is the pair correlation function of the gas, and \(S(r)\) is the Lorentz-Lekner screening factor.  By assuming for a structureless fluid~\cite{Lek67}
	\begin{equation*}
		g(r)=\begin{cases}
			0 \qquad \mbox{for }r<R_\mathrm{WS}\\
			1\qquad \mbox{for }r\ge R_\mathrm{WS}
		\end{cases}
	\end{equation*}
	and   \[S(r)=\left(1+\frac{8\pi}{3}\alpha N\right)^{-1}\]
\eqnref{eq:hernUP} gives
	\begin{equation}
		U_P(N)= -\frac{B}{2}\frac{\alpha e}{4\pi\epsilon_0}\left[\frac{N^{4/3}}{1+\frac{8\pi}{3\alpha }N}\right]
		\label{eq:UP}
	\end{equation}
	in which \(B\approx 20\) is a constant~\cite{Fermi1934,Her91,Spri68}. \eqnref{eq:UP} should be quite accurate at high density, where~\eqnref{eq:evko} is not.
	Thus, we have circumvented the difficulty of computing \(E_k\) at high density in the following way. We invert~\eqnref{eq:v0} so as to get
	\[E_k(N)= V_0(N) -U_P(N)\] 
	We used the experimental determination of \(V_0\) in argon~\cite{Rei83} (corrected for the offset due to impurity condensation on the electrodes~\cite{Bor91}) and subtracted \(U_P,\)~\eqnref{eq:UP}, from it. The resulting \(E_k(N)\) is displayed as a solid line in~\figref{fig:figure18}. The agreement with the experimentally determined \(E_k(N)\) values is remarkable.
	
\begin{figure}[b!]
		\centering
		\includegraphics[width=\columnwidth]{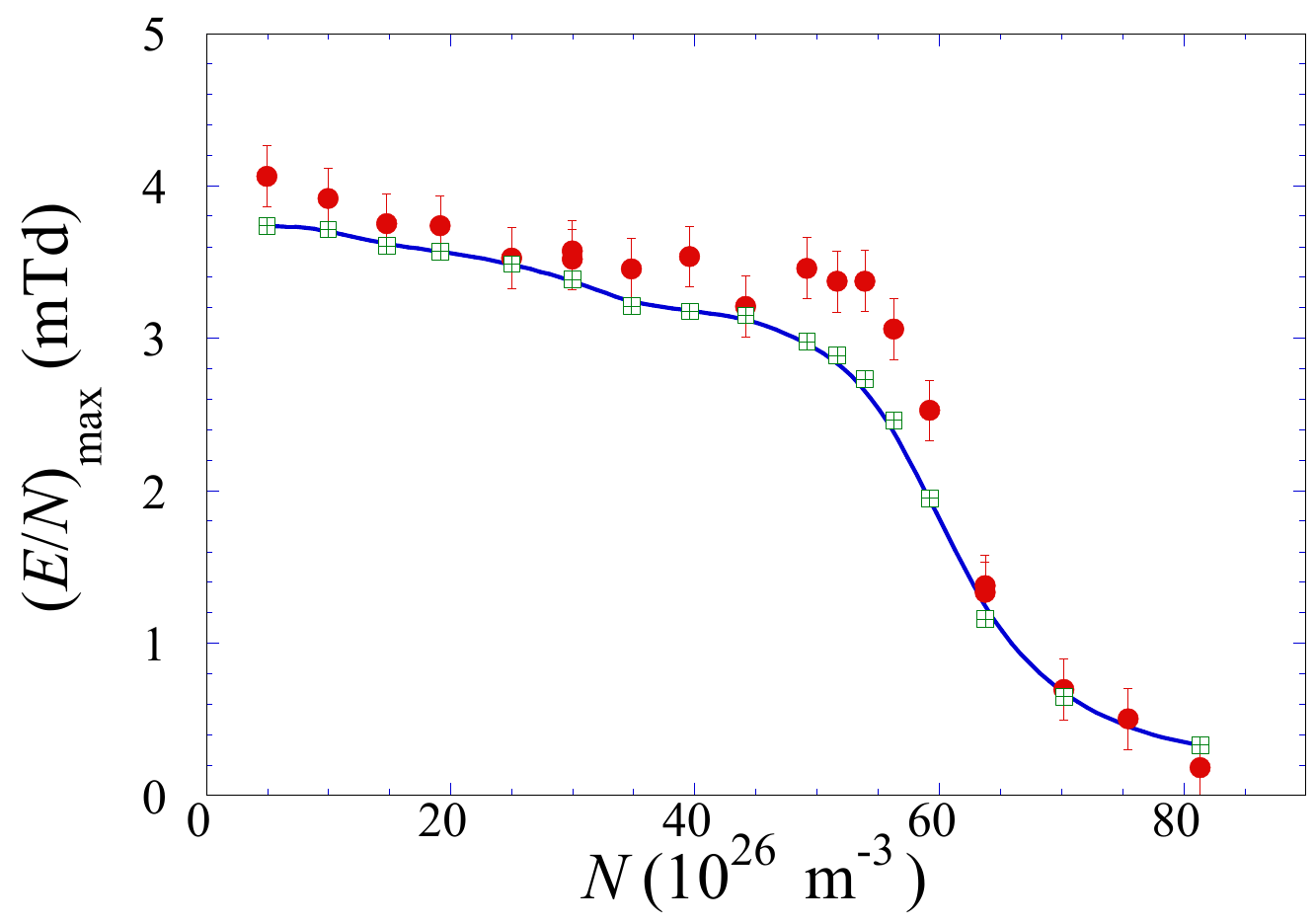}
		\caption{\small Density dependence of the reduced electric field \((E/N)_\mathrm{max}\) at the mobility maximum for  \(T=152.7\,\)K. Circles: experimentally determined values. Squares: values computed according the heuristic model by using the kinetic energy shift reported in~\figref{fig:figure18}. The solid line is only a guide for the eye.\label{fig:figure19}}
	\end{figure}

	At the same time, we display in~\figref{fig:figure19}
	the experimental values of \((E/N)_\mathrm{max}\) and compare them with those computed by using the \(E_k\) data shown in~\figref{fig:figure18}.  Taking into account the uncertainty in the available cross sections, we believe that the agreement of theory and experiment is rather good.  
	
	We note that \((E/N)_\mathrm{max}\) practically vanishes for \(N\gtrsim 80\times 10^{26}\,\)m\(^{-3}\). The reason is that the 
	kinetic energy shift is now so large that the maximum of the electron energy distribution function \(g(\varepsilon)\) falls beyond \(\varepsilon_\mathrm{RT}\)  and the peculiar features of the cross section are washed away by the integration in~\eqnref{eq:muN}.
	
	In a previous experiment~\cite{Bor2001}, carried out at \(T=152.15\,\)K, i.e., closer to \(T_c\), we were able to reach 
	the even higher density \(N\approx 140\times 10^{26}\,\)m\(^{-3}\).  In this case, \(\mu_0 N\) shows a maximum \(\mu_0 N\approx 28.6\times 10^{26}\,\)V\(^{-1}\,\)m\(^{-1}\,\)s\(^{-1}\) at exactly the same density \(N\approx 125.2\times 10^{26}\,\)m\(^{-3}\) as it has been observed in liquid argon at coexistence~\cite{Lam94}.  At such high densities, i.e., for \(N> N_M,\) the heuristic model is no longer able to reproduce the density dependence of \(\mu_0 N.\) However, by introducing an {\em ad hoc} density-dependent coefficient \(c_0(N)\) such that the effective scattering cross section is \(\sigma_\mathrm{eff}= c_0(N)\sigma_\mathrm{mt}^\star\), we were able to describe the reduced field dependence of \(\mu N\) and the location of the maximum of \(\mu_0 N \)
	at such high densities~\cite{Bor2001}.  As the mobility maximum in the liquid was explained in terms of the deformation potential theory by invoking phonon scattering~\cite{Bas79,Asc86,Nav93}, we draw the conclusion that for such high densities the {\em atomic} picture of the CKT, although heuristically modified by the incorporation of the MSE, breaks down, and an alternative approach has to be sought.

	\section{Conclusions}\label{sec:conc}
	
	This set of measurements completes our efforts to find a unified description of low-energy electron-atom scattering processes in high-density noble gases and some molecular ones. 
	
	Over the years, our research program has addressed helium, neon, argon, and, recently, hydrogen. In all of these gases, we have identified and highlighted three main multiple scattering effects that are heuristically incorporated into the classical kinetic theory for electron scattering in a low-density gas. 
	
	The resulting model does not rely on any adjustable parameters as it requires only the electron–atom momentum-transfer scattering cross section and the equation of state of the gas under investigation as input.
	
	One possible way to interpret the equations of the heuristic model is to recognize that an effective, density-dependent cross section is introduced into the framework of classical kinetic theory while retaining the picture of binary collisions. In this respect, the model is reminiscent of the Boltzmann-equation approach to describe electron drift velocity in liquids~\cite{Coh67,Sak86}. In that context as well, effective cross sections are introduced in the Boltzmann method.
	
	The present heuristic model is very successful in rationalizing the experimental results in wide density, temperature, and electric field ranges, as it has been shown in this paper for argon but also in the other gases we have investigated. However, it is clearly dependent on a good knowledge of the cross section, especially at low energies, where it is not always measured to a high (and sufficient) accuracy.
	
	The heuristic model is intrinsically subject to a limitation related to the gas density. In gases such as helium, neon, and hydrogen, where the low-energy electron–atom interaction is dominated by repulsive, short-range exchange forces, a different phenomenon emerges at sufficiently high densities, namely electron self-localization in cavities (for details, see Ref.~\cite{Borghesani2007}). In this regime, electrons are no longer quasi-free, and the electron drift velocity decreases by several orders of magnitude compared to that in the gas phase. Under such conditions, the binary-collision approach becomes meaningless, and electron transport must be treated within a hydrodynamic framework in the continuum limit. 
	
	Such a localization phenomenon has never been observed in argon. However, in both liquid argon and dense argon gas, a mobility maximum is observed at high density, followed by a decrease. It would therefore be interesting to extend the measurements to even higher gas densities to investigate the possibility that electrons might self-localize in higher-than-average density fluctuations or in clusters
	 due to the negative values of their energy \(V_0\)
	at the bottom of the conduction band, even though we expect that the necessary large density values might not be reached under usual conditions.
	
	\begin{acknowledgments}
		The authors gratefully acknowledge enlightening discussions with late Prof. M. Santini and also 
		dr. D. Neri.
	\end{acknowledgments}
	\subsection*{Conflict of interest}
	The authors have no conflict of interest.
	\subsection*{Authors contribution}
	All authors contributed equally.
	\subsection*{Data availability}
	The data that support the findings of this study are available from
	the corresponding author upon  request.
	 \section*{References}
	\bibliography{eArgon.bib}
\end{document}